\documentclass[twocolumn,amsmath,floatfix,amssymb,aps,prb,showpacs,superscriptaddress]{revtex4-1}
\usepackage{graphicx,natbib}
\usepackage{color}

\begin{document}

\title{Neutron spectroscopic study of crystal field excitations in Tb$_2$Ti$_2$O$_7$ and Tb$_2$Sn$_2$O$_7$}

\author{J.~Zhang}
\affiliation{Department of Physics and Astronomy, McMaster University, Hamilton, Ontario, L8S 4M1, Canada}
\author{K.~Fritsch}
\affiliation{Department of Physics and Astronomy, McMaster University, Hamilton, Ontario, L8S 4M1, Canada}
\author{Z.~Hao}
\affiliation{Department of Physics and Astronomy, University of Waterloo, Waterloo, Ontario, N2L 3G1, Canada}
\author{B.~V.~Bagheri}
\affiliation{Department of Physics and Astronomy, University of Waterloo, Waterloo, Ontario, N2L 3G1, Canada}
\author{M.~J.~P.~Gingras}
\affiliation{Department of Physics and Astronomy, University of Waterloo, Waterloo, Ontario, N2L 3G1, Canada}
\affiliation{Canadian Institute for Advanced Research, 180 Dundas St. W., Toronto, Ontario, M5G 1Z8, Canada}
\affiliation{Perimeter Institute of Theoretical Physics, 31 Caroline North, Waterloo, Ontario, N2L 2Y5, Canada}
\author{G.~E.~Granroth}
\affiliation{Quantum Condensed Matter Division, Oak Ridge National Laboratory, Oak Ridge, Tennessee, 37831, USA}
\author{P.~Jiramongkolchai}
\affiliation{Department of Chemistry, Princeton University, Princeton, New Jersey, 08544, USA}
\author{R.~J.~Cava}
\affiliation{Department of Chemistry, Princeton University, Princeton, New Jersey, 08544, USA}
\author{B.~D.~Gaulin}
\affiliation{Department of Physics and Astronomy, McMaster University, Hamilton, Ontario, L8S 4M1, Canada}
\affiliation{Canadian Institute for Advanced Research, 180 Dundas St. W., Toronto, Ontario, M5G 1Z8, Canada}
\affiliation{Brockhouse Institute for Materials Research, McMaster University, Hamilton, Ontario, L8S 4M1, Canada}

\date{14 April 2014}

\begin{abstract}
We present time-of-flight inelastic neutron scattering measurements at low temperature on powder samples of the magnetic pyrochlore oxides Tb$_2$Ti$_2$O$_7$ and Tb$_2$Sn$_2$O$_7$. These two materials possess related, but different ground states, with Tb$_2$Sn$_2$O$_7$ displaying ``soft" spin ice order below $T_{\rm N}\sim 0.87$ K, while Tb$_2$Ti$_2$O$_7$ enters a hybrid, glassy spin ice state below $T_{\rm g}\sim 0.2$ K. Our neutron measurements, performed at $T=1.5$ K and $30$ K, probe the crystal field states associated with the $J$ = 6 states of Tb$^{3+}$ within the appropriate $Fd\bar{3}m$ pyrochlore environment. These crystal field states determine the size and anisotropy of the Tb$^{3+}$ magnetic moment in each material's ground state, information that is an essential starting point for any description of the low-temperature phase behavior and spin dynamics in Tb$_2$Ti$_2$O$_7$ and Tb$_2$Sn$_2$O$_7$. While these two materials have much in common, the cubic stanate lattice is expanded compared to the cubic titanate lattice. As our measurements show, this translates into a factor of $\sim~2$ increase in the crystal field bandwidth of the $2J+1=13$ states in Tb$_2$Ti$_2$O$_7$ compared with Tb$_2$Sn$_2$O$_7$. Our results are consistent with previous measurements on crystal field states in Tb$_2$Sn$_2$O$_7$, wherein the ground-state doublet corresponds primarily to $m_J=\left|\pm 5\right\rangle$ and the first excited state doublet to $m_J=\left|\pm 4\right\rangle$. In contrast, our results on Tb$_2$Ti$_2$O$_7$ differ markedly from earlier studies, showing that the ground-state doublet corresponds to a significant mixture of $m_J=\left|\pm 5\right\rangle$, $\left|\mp 4\right\rangle$, and $\left|\pm 2\right\rangle$, while the first excited state doublet corresponds to a mixture of $m_J=\left|\pm 4\right\rangle$, $\left|\mp 5\right\rangle$, and $\left|\pm 1\right\rangle$. We discuss these results in the context of proposed mechanisms for the failure of Tb$_2$Ti$_2$O$_7$ to develop conventional long-range order down to 50 mK.

\end{abstract}


\maketitle

\section{Introduction}

Geometrically frustrated magnetic materials combine magnetic moments with interactions and anisotropies that inhabit certain crystalline architectures such that a conventional ordered state cannot easily form at low temperatures.\cite{Lacroix2011} In two dimensions, the best appreciated combinations include simple antiferromagnetism and two-dimensional assemblies of triangles, as occurs in either the triangular \cite{Kadowaki1987,Collins1997} or kagome lattice antiferromagnets. \cite{Inami2000,Wills2000,Han}  In three dimensions, the combination of local Ising anisotropy and net ferromagnetic interactions on the pyrochlore lattice, a three-dimensional network of corner-sharing tetrahedra, leads to spin ice physics.\cite{Ramirez1999,Bramwell2001,Harris1997}  Magnetism in materials based on such a pyrochlore lattice structure often leads to exotic ground states which may lack long-range order, such as spin glass,\cite{GardnerGaulin1999,Singh2012,Singh2008,Greedan1996,Gaulin1992,Greedan1991,Gingras1997} spin liquid \cite{Kanada1999,Mirebeau2007,Gardner1999,Gardner2001,Gardner2003} and spin ice state,\cite{Ramirez1999,Hertog2000,Harris1997,Bramwell2001,Ross2011,Kimura2013} as well as ordered magnetic structures that are selected by exotic mechanisms such as quantum or thermal order by disorder.\cite{Champion2003,Savary2012,Zhitomirsky2012}

A great variety of cubic pyrochlore magnets with chemical formula $R_2M_2O_7$ crystallizes in the $Fd\bar{3}m$ space group.  In this formula, $R$ is a magnetic trivalent rare-earth ion (e.g. Gd, Tb, Dy, Ho, Er and Yb) with eightfold oxygen coordination (Fig. 1) and $M$ can be either a nonmagnetic (Ti, Zr, Sn) or a magnetic (Mn, Mo) tetravalent transition-metal ion with sixfold oxygen coordination.\cite{Gardner2010} The two sublattices of $R$ ions and $M$ ions form independent and interpenetrating networks of corner-sharing tetrahedra.\cite{Gardner2010} Much attention has focused on those cubic pyrochlores for which only one of the $R$ or $M$ sublattices is magnetic (see, however, Refs. [11,12]) as this tends to distill out the physics of interacting magnetic moments on a simpler, three-dimensional, frustrated crystalline architecture.  

The ground states exhibited by the rare earth pyrochlores are varied due to the breadth of magnetic interactions and anisotropies that they display. 
The angular momentum $J$ of a free $R^{3+}$ ion is determined by applying Hund's rule to its partially filled $4f$ shell.  In $R_2M_2O_7$, each $R^{3+}$ ion is surrounded by eight oxygen ions. The local electric potential environment created by these oxygen ions at the $R^{3+}$ site, the ``crystal field",  lowers the full O(3) rotation symmetry to the symmetry of a cube compressed along one of its body diagonals, $D_{3d}$. As a result, the $2J+1$ levels split into the energy eigenstates of the crystal field Hamiltonian. The energies and wave functions of these crystal field levels determine the size and anisotropy of the magnetic moments of the $R^{3+}$ ions, as well as their interactions at low temperature.  

Understanding the size, anisotropy, and ultimately the wave functions associated with the ground state magnetic moment in the $R_2M_2O_7$ pyrochlores is an essential starting point to the full description of their exotic low-temperature properties.  Estimates for the crystal field eigenfunctions and eigenvalues for many $R^{3+}$ ions in $R_2M_2O_7$ pyrochlores have been derived from inelastic neutron scattering.\cite{Mirebeau2007,Zinkin1996,Rosenkranz2000,Champion2003,Gardner2009,Bertin2012,Gingras2000,Princep2013} As the crystal field eigenvalues in these systems can extend to energies of $\sim~100$~meV, the unique determination of the eigenvalues and eigenfunctions of the crystal field levels is not necessarily straightforward.  Recent advances in time-of-flight neutron spectroscopy have greatly enhanced our capability to measure such crystal field (and other magnetic) excitations over a broad dynamic range in energy.  It is therefore timely to re-examine some of the earlier crystal field studies, especially in cases where the precise details of the crystal field eigenfunctions are believed to be important to ground-state selection and to the nature of the experimentally observed exotic ground state.

Important cases are the crystal field eigenfunctions and eigenvalues for Tb$^{3+}$ in Tb$_2$Ti$_2$O$_7$ and Tb$_2$Sn$_2$O$_7$.  These two isostructural compounds are both cubic pyrochlores with antiferromagnetic Curie-Weiss constants of $\Theta_{\rm CW}\sim-19$ K \cite{Gardner1999,Gardner2001,Mirebeau2007,Gingras2000} and $\sim-12$~K,\cite{Matsuhira2002} respectively (note that estimates of $\Theta_{\rm CW}$ with the contribution of the crystal field field states taken into account result in smaller, but still antiferromagnetic, $\Theta_{\rm CW}$ values \cite{Mirebeau2007,Gingras2000}). In both materials, the $M$ site is nonmagnetic, populated by Ti$^{4+}$ or Sn$^{4+}$, respectively. Tb$_2$Sn$_2$O$_7$ is strongly ($\sim~3\%$) expanded compared with Tb$_2$Ti$_2$O$_7$ with room temperature lattice parameters of 10.426~\AA \cite{Mirebeau2005} and 10.149~\AA, \cite{Gardner2001} respectively. In both materials, the low-lying crystal field excitations have been previously studied using triple axis neutron scattering techniques.\cite{Mirebeau2007,Gingras2000} The picture of a ground-state doublet separated by $\sim 1.5$~meV from an excited state doublet in both materials has been known for some time.\cite{Gardner2001,Mirebeau2007,Gingras2000} The next highest energy eigenstates begin at $\sim10$~meV.\cite{Gardner2001,Mirebeau2007,Gingras2000} The nature of the crystal field eigenstates at energies large compared to $\sim10$~meV, where at least half of the $2J+1=13$ crystal field states reside, are less well understood.  Moreover, the precise details of even the low-lying crystal field states, which determine the ground-state and low-temperature properties, are overall better understood in the context of a more comprehensive study extending over a broader dynamic range in energy, as allowed by time-of-flight neutron techniques.

The nature of the ground state in Tb$_2$Ti$_2$O$_7$ has been debated for almost 15 years, since it was realized that it does not display a conventional long-range ordering down to $\sim0.05$~K, \cite{Gardner1999,Gardner2001} in spite of its $\Theta_{\rm CW}\sim-19$~K. It does show evidence for a magnetically frozen state at temperatures below $\sim0.2$~K, \cite{Gardner2003,Hamaguchi2004,Yasui2002,Lhotel2012} and there has been a flurry of recent neutron scattering work, showing evidence for both ``pinch point" diffuse scattering and short-range frozen order with a ($\frac{1}{2},\frac{1}{2},\frac{1}{2}$) ordering wavevector, also at temperatures of $\sim0.2$~K. \cite{Fritsch2013,Fennell2012,Petit2012,Fritscharxiv}  The diffuse Bragg-like peaks observed at the ($\frac{1}{2},\frac{1}{2},\frac{1}{2}$) ordering wavevector have been modeled on the basis of an antiferromagnetically correlated ``soft" spin ice state, where all spins within the cubic unit cell point either into or out of the tetrahedra on which they reside, locally satisfying the two-spins-in$-$two-spins-out  ``ice rules".\cite{Fritsch2013}  Neighboring cubic unit cells are $\pi$ out of phase with respect to each other, and the resulting order extends only over a couple of unit cells. The ``soft" descriptor here refers to the fact that the spins do not point exactly along local $\left\langle111\right\rangle$ directions, that is, they do not point exactly into or out of the tetrahedra, but make an angle of $\sim10^{\circ}$ with respect to their local $\left\langle111\right\rangle$ directions. In addition, recent work\cite{Taniguchi2013} on polycrystalline samples of Tb$_{2+x}$Ti$_{2-x}$O$_{7+y}$, with $x$ and $y$ within 0.5$\%$ of stoichiometry, have shown an anomaly in the heat capacity near 0.4 - 0.5 K which is very sensitive to the exact value of $x$ and $y$.

Two theoretical proposals have been put forward as to why Tb$_2$Ti$_2$O$_7$ fails to order at the ``expected" temperatures $\sim$ 1 K,\cite{Hertog2000} both of which critically involve the crystal field eigenfunctions and eigenvalues for Tb$^{3+}$ in Tb$_2$Ti$_2$O$_7$.  
One of these, referred to as ``quantum spin ice", \cite{Molavian2007} 
proposes that interaction-induced virtual transitions between the ground-state doublet and first excited state crystal field doublet of Tb$^{3+}$ renormalize the antiferromagnetic effective exchange interaction to ferromagnetic.
This leads to a spin ice state,  akin to that displayed by the ``classical" spin ice systems Ho$_2$Ti$_2$O$_7$ \cite{Harris1997,Fennell2009,Clancy2009} and Dy$_2$Ti$_2$O$_7$,\cite{Ramirez1999}
but now with the addition of quantum spin fluctuations \cite{Hermele,Savary,SBLee}
that are concurrently generated through the virtual transitions between the crystal field states.~\cite{Molavian2007} 
These virtual transitions are likely quite strong in Tb$_2$Ti$_2$O$_7$ 
due to the proximity in energy of the ground-state and first excited state doublets. 
The competing theoretical scenario involves the fact that a Tb$^{3+}$ ion, in its $4f^8$ configuration, possesses an even number of electrons. The doublet nature of its crystal field ground state is therefore not protected by Kramers' theorem. Bonville and collaborators have proposed that Tb$_2$Ti$_2$O$_7$ undergoes a Jahn-Teller phase transition at low but finite temperature to a singlet ground state, creating a sufficiently large singlet-singlet gap such that dipolar magnetic order does not occur.\cite{Bonville2011} While this is an attractive scenario in several respects, no direct evidence exists for a lowering of the crystal symmetry of Tb$_2$Ti$_2$O$_7$ below cubic, \cite{Gaulin2011} 
despite serious attempts to look for the relevant splitting of cubic Bragg peaks. \cite{Ruff2007} 
Strong magnetoelastic effects have been observed in dilatometry and Young's modulus measurements, \cite{Mamsurova1986, Nakanishi2011} and high resolution x-ray scattering down to 0.3 K and in zero field show evidence for Jahn-Teller-like fluctuations, \cite{Ruff2007} but no splitting is observed except under the application of very high magnetic fields ($\sim 25$ T). \cite{Ruff2010}  In the absence of such a splitting, the evidence for the hypothetical Jahn-Teller distortion comes principally from intensities of crystal field transitions out of the ground state.\cite{Bonville2011}

In contrast to Tb$_2$Ti$_2$O$_7$, Tb$_2$Sn$_2$O$_7$ can only be studied in polycrystalline form, due to the strong tendency of SnO$_2$ to sublimate at modest temperatures. While less extensively studied than Tb$_2$Ti$_2$O$_7$ for that reason, it has also been less controversial in part because it does exhibit clear signs of magnetic order at temperatures of $\sim1$~K.  The expanded lattice of Tb$_2$Sn$_2$O$_7$ compared with Tb$_2$Ti$_2$O$_7$ tends to strengthen ferromagnetic near-neighbor dipolar interactions relative to antiferromagnetic exchange. This tendency appears to be strong enough in Tb$_2$Sn$_2$O$_7$ that it displays a two-step ordering process with ferromagnetic correlations appearing rapidly below $\sim1.3$~K, soon followed by a phase transition to an ordered ``soft" spin ice state below $T_{\rm C}=0.87$~K. \cite{Mirebeau2005,Reotier2006} Again, the ``soft" descriptor refers to the fact that the Tb$^{3+}$ moments are oriented at an angle of $13.3^\circ$ with respect to their local $\left\langle111\right\rangle$ directions (directly into or out of the tetrahedra). \cite{Mirebeau2005,Rule2009} The two-spins-in$-$two-spins-out ``ice rules" of the local [111] projection of the magnetic moments are obeyed within the soft ordered spin ice structure of  Tb$_2$Sn$_2$O$_7$ below $T_{\rm C}$. It is an ordered state, distinct from the disordered ``classical" spin ice state by virtue of the long-range correlations present within it. Despite the evidence provided for long-range order by neutron scattering, $\mu$SR measurements show this ordered spin ice state to coexist with slow, correlated magnetic fluctuations on the time scale of 10$^{-4}$~s to 10$^{-8}$~s and which persist down to the lowest temperatures measured. \cite{Reotier2006,Bert2006}  

The above discussion makes it rather clear that, independent of the nature of the physics operating in Tb$_2$Ti$_2$O$_7$ that endows it with complex low-temperature 
($T\lesssim0.5$ K) properties and properties different from its sister compound Tb$_2$Sn$_2$O$_7$, this physics must depend on the nature of the crystal electrical field states, and in particular their $\vert m_j\rangle $ spectral decomposition.
Indeed, both the quantum spin ice and the split-doublet theoretical scenarios depend on the specific nature of that decomposition.
Ideally, for the quantum spin ice scenario leveraged from virtual crystal field fluctuations, one would need to have access to the bare (unrenormalized) wave
functions of the isolated Tb$^{3+}$ ions in order to construct the effective low-energy theory of the material. \cite{Molavian2007,Molavian_arXiv,McClarty2010}
As a first step towards this goal, we reinvestigate the crystal field states of ``dense'' 
Tb$_2$Ti$_2$O$_7$ and Tb$_2$Sn$_2$O$_7$ to ascertain that, at least, the ``dressed states'' (dressed by interactions) are well characterized.
The presence of interactions leads to an entanglement of the crystal field wave functions of the otherwise free (noninteracting) Tb$^{3+}$.
How to describe these dressed states in the presence of strong correlations when the bare states are not known is a difficult problem.
In the present work, we sidestep this difficulty by parametrizing the transitions we determined via inelastic neutron scattering using a simple
noninteracting single-ion crystal field Hamiltonian. Practically, this amounts to neglecting the dispersion of the crystal field levels generated
by the interactions. As we see below, only one set of measured transitions in both Tb$_2$Ti$_2$O$_7$  and Tb$_2$Sn$_2$O$_7$ displays measurable dispersion.  This is relatively weak, and does not persist to the $T=30$~K data. We thus consider that proceeding with a single-ion Hamiltonian to parametrize the crystal field transitions observed in 
Tb$_2$Ti$_2$O$_7$ and Tb$_2$Sn$_2$O$_7$ is a reasonable starting point.

In this paper, we present new time-of-flight inelastic neutron scattering measurements on polycrystalline samples of Tb$_2$Ti$_2$O$_7$ and Tb$_2$Sn$_2$O$_7$ at energy transfers less than $\sim90$ meV. These measurements were performed at two temperatures, 1.5 K and 30 K, such that we can observe transitions both out of the ground-state doublet alone, and out of the combination of both the ground-state and first excited state crystal field state doublets.  Parameters describing the crystal fields appropriate to both Tb$_2$Ti$_2$O$_7$ and Tb$_2$Sn$_2$O$_7$ are then determined by fitting these measurements to the results of crystal field calculations, and the eigenfunctions and eigenvalues appropriate to Tb$^{3+}$ in both environments are determined.  In Sec. \textrm{II}, we present the details of the sample preparation and the neutron scattering measurements.  We then introduce in Sec. \textrm{III} the theoretical calculations of the crystal field Hamiltonian and the magnetic neutron scattering cross section. In Sec. \textrm{IV}, we compare the neutron scattering results and appropriate calculations and describe the resulting crystal field eigenvalues and eigenfunctions for Tb$_2$Ti$_2$O$_7$ and Tb$_2$Sn$_2$O$_7$.  These results are discussed in Sec. \textrm{V}.

\section{EXPERIMENTAL DETAILS}

Powder samples of Tb$_2$Ti$_2$O$_7$ and Tb$_2$Sn$_2$O$_7$ were synthesized by mixing high purity Tb$_4$O$_7$ with TiO$_2$ and SnO$_2$, respectively, in the appropriate stoichiometric ratios, and placing these in high density aluminum oxide crucibles. The materials were fired in air for two days at temperatures of 1100 $^\circ$C, 1200 $^\circ$C, 1300 $^\circ$C and finally 1350 $^\circ$C, with intermediate grindings. Powder x-ray diffraction measurements showed high-quality single-phase materials displaying the cubic space group $Fd\bar{3}m$.

Inelastic neutron scattering measurements were performed on the SEQUOIA direct geometry time-of-flight spectrometer \cite{Granroth2010} at the Spallation Neutron Source (SNS) of Oak Ridge National Laboratory.  Measurements were performed at $T=1.5$~K and 30~K and over a wide dynamic range of energy transfer, using incident neutron energies, $E_i$, of 11, 45 and 120 meV, in order to probe the crystal field excitations over a large range in energy.  Measurements employing higher incident energies cover a larger dynamic range, and correspond to coarser energy transfer resolution, which is typically 2-3$\%$ of $E_i$.

Two 100 mm diameter Fermi choppers were used for these measurements: the coarse-resolution chopper had a 3.5 mm slit spacing and the fine-resolution chopper had a 2 mm slit spacing.  For  $E_i$=120 meV, the coarse resolution chopper was spun at 300 Hz  and the bandwidth-limiting $T_0$ chopper was spun at 180 Hz.  The fine resolution chopper was spun at 180 and 420 Hz for $E_i$ =11 and 45 meV, respectively.  For $E_i$=11 meV, the $T_0$ chopper operated at 60 Hz and, for $E_i$=45 meV, it operated at 90 Hz. In all cases, the corresponding energy resolution at the elastic position was $\sim3\%$ of $E_i$; 0.3 meV at $E_i$=11 meV, 1.3 meV at $E_i$=45 meV and 3.5 meV at $E_i$=120 meV.

The powder samples of Tb$_2$Ti$_2$O$_7$ and Tb$_2$Sn$_2$O$_7$ were placed in a flat plate Al cell to completely cover the 5~cm~$\times$~5~cm beam, and the Al cell was mounted in an Orange ILL cryostat with a base temperature of 1.5 K. The data were normalized to vanadium to remove variation in detector efficiency. Background data from an empty aluminum can was subtracted from the signal.

\section{CRYSTAL FIELD AND MAGNETIC NEUTRON SCATTERING CALCULATION}

\subsection{Crystal Field Calculations}

The Tb$^{3+}$ ion has eight electrons in the $4f$ shell with angular momentum $J=6$ according to Hund's rule. In Tb$_2M_2$O$_7$ ($M$=Ti, Sn), each Tb$^{3+}$ ion is surrounded by eight oxygen ions, which form a distorted cube (Fig. 1), compressed along one of its body diagonals. The crystal electric field created by these oxygen ions at the Tb$^{3+}$ site has $D_{3d}$ symmetry. As a result,  the crystal field Hamiltonian is invariant under three-fold rotations around the local [111] direction as well as point inversion with respect to the Tb$^{3+}$ site. Taking into account these constraints, the crystal field Hamiltonian can be written in terms of Steven's operator equivalents as\cite{Mirebeau2007,Gingras2000} 

\begin{equation}
\begin{split}
H_{\rm{cf}}=&\alpha _JD_2^0O_2^0+\beta _J\left( D_4^0O_4^0+D_4^3O_4^3\right)\\
&+\gamma _J(D_6^0O_6^0+D_6^3O_6^3+D_6^6O_6^6)
\end{split}
\end{equation}
where the coefficients are $\alpha_J=-1/99$, $\beta_J=2/16335$, $\gamma_J=-1/891891$ for Tb$^{3+}$.\cite{Hutchings} $D_n^m$ are the crystal field parameters that will be determined for Tb$_2M_2$O$_7$ with $M$=Ti and Sn in this paper. Another equivalent form for this Hamiltonian, expressed in terms of spherical harmonics $Y_n^m$ is:

\begin{equation}
           H'_{\rm{cf}}=\sum_{nm}B_n^m\sqrt{\frac{4\pi }{2n+1}}Y_n^m
\end{equation}
The two sets of coefficients, $D_n^m$ and $B_n^m$, in Eqs. (1) and (2) can be transformed into each other by $D_n^m=\lambda_n^mB_n^m$, where the $\lambda_n^m$ are given for Tb$^{3+}$ in Table $\textrm{I}$. \cite{Kassman1970}

\begin{figure}[h]
\includegraphics[width=7.5cm]{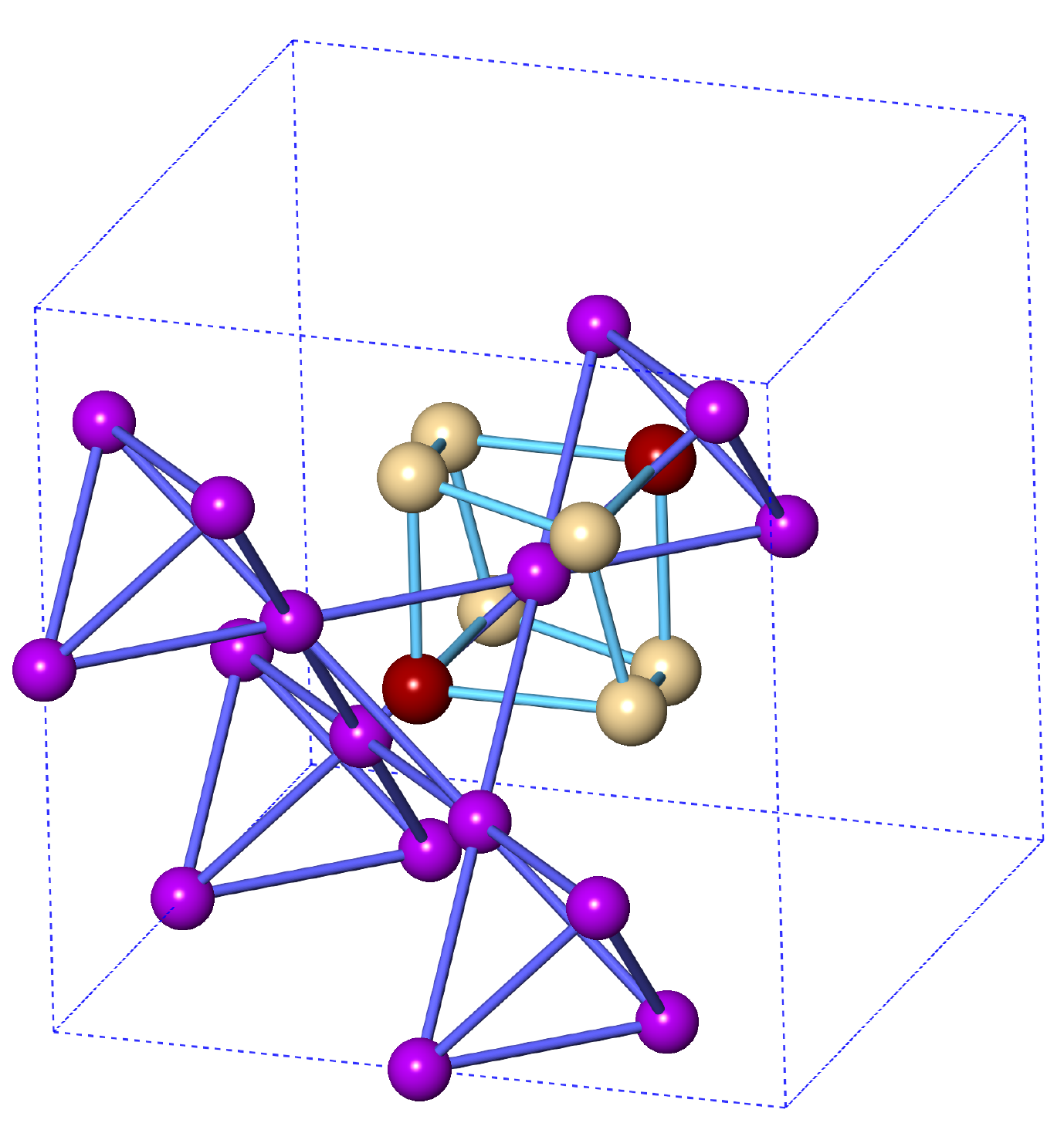}
\caption{(Color online) The crystal environment immediately around the Tb$^{3+}$ ion (purple) within Tb$_2M_2$O$_7$ is shown. Each Tb$^{3+}$ ions is surrounded by eight oxygens (red and
 beige at the vertices of the cube) which form a distorted cube. The two apical oxygens (red) along one diagonal of the cube (the local $\left[111\right]$ direction) are closer to the Tb$^{3+}$ ion than are the other six oxygens (shown in beige).}
\end{figure}

The crystal field parameters are difficult to calculate from first principles, so they are usually determined by comparison between theoretical crystal field calculations and experimental data. The most direct comparison is to inelastic neutron scattering data, as this technique directly probes the spectroscopy of the crystal field eigenfunctions, according to magnetic dipole selection rules.  We therefore turn to the calculation of the magnetic neutron scattering cross section relevant to such crystal field states.

\begin{table}[h]
\caption{Values of $\lambda_n^m$ for Tb$^{3+}$. \cite{Kassman1970}}
\begin{tabular*}{8.7cm}{{@{\extracolsep{\fill}}|cccccc|}} 
\hline \hline
$\lambda_2^0$&$\lambda_4^0$&$\lambda_4^3$&$\lambda_6^0$&$\lambda_6^3$&$\lambda_6^6$\\
\hline 
1/2&1/8&$-\sqrt{35}/2$&1/16&$-\sqrt{105}/8$&$\sqrt{231}/16$\\
\hline\hline
\end{tabular*}
\end{table}

\subsection{Magnetic Neutron Scattering Cross Section}

Following the notation of Squires,\cite{Squires1978} the magnetic neutron scattering cross section is given by the expression
\begin{equation}
\begin{split}
\frac{d^2\sigma }{d\Omega dE'}=&\left(\gamma r_0\right)^2\frac{k'}{k}\sum_{\alpha \beta }\left(\delta _{\alpha \beta} -\hat{\kappa} _\alpha \hat{\kappa} _\beta \right)\sum_{\lambda \lambda '}p_\lambda\\
 &\times\left\langle\lambda \left|Q_\alpha ^\dagger \right|\lambda '\right\rangle\left\langle\lambda '\left|Q_\beta \right|\lambda \right\rangle\delta \left( E_\lambda -E_{\lambda'}+\hbar\omega \right),
\end{split}
\end{equation}
where $Q_\alpha$ in Eq. (3) is related to the Fourier transform of the magnetization, and this expression can be simplified to
\begin{equation}
\frac{d^2\sigma }{d\Omega dE'}=C\frac{k'}{k}\sum_{\alpha}\sum_{\lambda \lambda '}p_\lambda \left\langle\lambda \left|J_\alpha ^\dagger \right|\lambda '\right\rangle\left\langle\lambda '\left|J_\alpha \right|\lambda \right\rangle L_{\lambda \lambda '}
\end{equation}
where $C$ is a constant, $k$ and $k'$ are the moduli of the incident and scattered wavevectors. $\sum_{\alpha}\sum_{\lambda \lambda '}p_\lambda \left\langle\lambda \left|J_\alpha ^\dagger \right|\lambda '\right\rangle\left\langle\lambda '\left|J_\alpha \right|\lambda \right\rangle$ is the sum over the intensities of the transitions from crystal field state $\left|\lambda\right\rangle$ of energy $E_\lambda$ to the crystal field state $\left| \lambda'\right\rangle$ of energy $E_{\lambda'}$,  $p_\lambda$ is the probability that the crystal is initially in the state $\lambda$ and $J_\alpha$ is the $x$, $y$, or $z$ component of the total angular momentum operator. 

$L_{\lambda\lambda'}$ is a Lorentzian function, which describes the lineshape of the transition from state $\left|\lambda\right\rangle$ of energy $E_\lambda$ to the state $\left|\lambda'\right\rangle$ of energy $E_{\lambda'}$:
\begin{equation}
L_{\lambda \lambda '}=\frac{1}{\pi }\frac{\Gamma _{\lambda \lambda '}}{\Gamma _{\lambda \lambda '}^2+\left[\hbar\omega -\left(E_{\lambda '}-E_\lambda \right)\right]^2},
\end{equation}
This function replaces the delta function in the energy difference between the two states in Eq. (3). As crystal field states typically have little dispersion and a long lifetime, the energy width of $L_{\lambda\lambda'}$,  $\Gamma _{\lambda \lambda '}$, is usually determined by the finite instrumental energy resolution of the spectrometer. As a rule, the lower the incident energy, the better the resolution, which means the narrower the relevant inelastic peak in the spectrum. 

\section{INELASTIC NEUTRON SCATTERING AND CRYSTAL FIELD CALCULATION RESULTS}

\begin{figure}[h]
\centering
\includegraphics[width=\columnwidth]{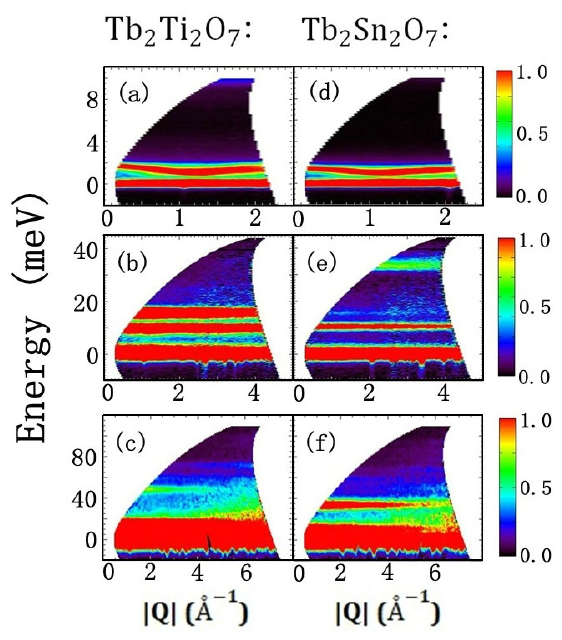}
\caption{(Color online) Color contour maps of energy vs $\vert {\bf Q} \vert$ from the inelastic neutron scattering data for Tb$_2$Ti$_2$O$_7$ and Tb$_2$Sn$_2$O$_7$ at $T=1.5$~K are shown. The top two plots show  data at relatively low energies taken with $E_i=11$ meV neutrons. Well-defined transitions between the ground-state crystal field doublet and the lowest excited state crystal field doublet between 1-2 meV as well as weak dispersion of this excitation, are observed for both samples. The middle panels show data up to $\sim43$ meV, using $E_i=45$ meV neutrons. The bottom panels show data taken to the highest energy transfers, employing $E_i=120$ meV neutrons. All data sets shown are for $T=1.5$~K. An empty can background was subtracted and the data were corrected for detector efficiency.}
\end{figure}

We show in Fig. 2 the inelastic neutron scattering data for Tb$_2$Ti$_2$O$_7$ (left side) and Tb$_2$Sn$_2$O$_7$ (right side) taken on the SEQUOIA time-of-flight chopper spectrometer at the SNS.\cite{Granroth2010}  Figures 2(a)-(c) show data for Tb$_2$Ti$_2$O$_7$ using (a) $E_i=11$, (b) 45, and (c) 120 meV, while Figs. 2(d)-(f) show the same $E_i=11$, 45, and 120 meV data sets, respectively, for Tb$_2$Sn$_2$O$_7$.  As nearly identical amounts of the two Tb$_2$Ti$_2$O$_7$ and Tb$_2$Sn$_2$O$_7$ powder samples were loaded into identical sample cans in the same cryostat, a direct comparison can be made between the two sets of data, all taken at $T=1.5$~K. These color contour maps show the full energy vs $\vert {\bf Q} \vert$ data sets.  Inspection of Fig. 2 shows a large number of excitations which are mostly identified as magnetic crystal field excitations, due to their weak dispersion and their $\vert {\bf Q} \vert$ independence.  The top data sets, in Fig. 2(a) for Tb$_2$Ti$_2$O$_7$ and Fig. 2(d) for Tb$_2$Sn$_2$O$_7$, show the relatively low-energy crystal field excitations below $\sim8$~meV, and show great similarity between the two pyrochlores. The lowest-lying crystal field excitations, between 1 and 2 meV in energy, do exhibit weak dispersion in these $T=1.5$~K data sets in both materials, as was previously reported. \cite{Gardner1999,Kanada1999,Mirebeau2007} This dispersion disappears at temperatures large compared to $\Theta_{\rm CW}$, and it is not observed in our $T=30$~K data sets. This dispersion develops inversely with the strength of the peak in the magnetic diffuse scattering, and is attributed to the formation of strong short-range spin correlations. A random phase approximation (RPA) calculation for a simple Hamiltonian for Tb$_2$Ti$_2$O$_7$ can explain this disappearance of the dispersion for $T>\Theta_{\rm CW}$.\cite{Kao2003} Clear differences between Tb$_2$Ti$_2$O$_7$ and Tb$_2$Sn$_2$O$_7$ are evident at intermediate energies, up to $\sim40$~meV as seen in Figs. 2(b) and 2(e), and these differences persist to the highest energies measured, $\sim90$~meV, as shown in Figs. 2(c) and 2(f).  

Consistent with previous measurements,\cite{Gardner1999,Kanada1999,Mirebeau2007} we see that the lowest energy crystal field excited states, at the minimum of their dispersion, in either Tb$_2$Ti$_2$O$_7$ or Tb$_2$Sn$_2$O$_7$ are at $\sim1$ meV $\sim11$~K, and therefore the base temperature of $T=1.5$~K assures us that we are observing transitions out of the ground state only. As the lowest-lying excited crystal field states are confined to energies less than 2 meV $\sim22$ K, and the next highest-energy crystal field states are at $\sim 10$ meV $\sim110$ K, our measurements at $T=30$~K correspond almost solely to the ground-state and first excited state crystal field excitations being populated. Thus, at 30 K, we observe transitions from both sets of these low-lying crystal field states, but not out of the excited levels at an energy $>10$ meV ($\sim110$ K).

\begin{figure}[h]
\includegraphics[width=8cm]{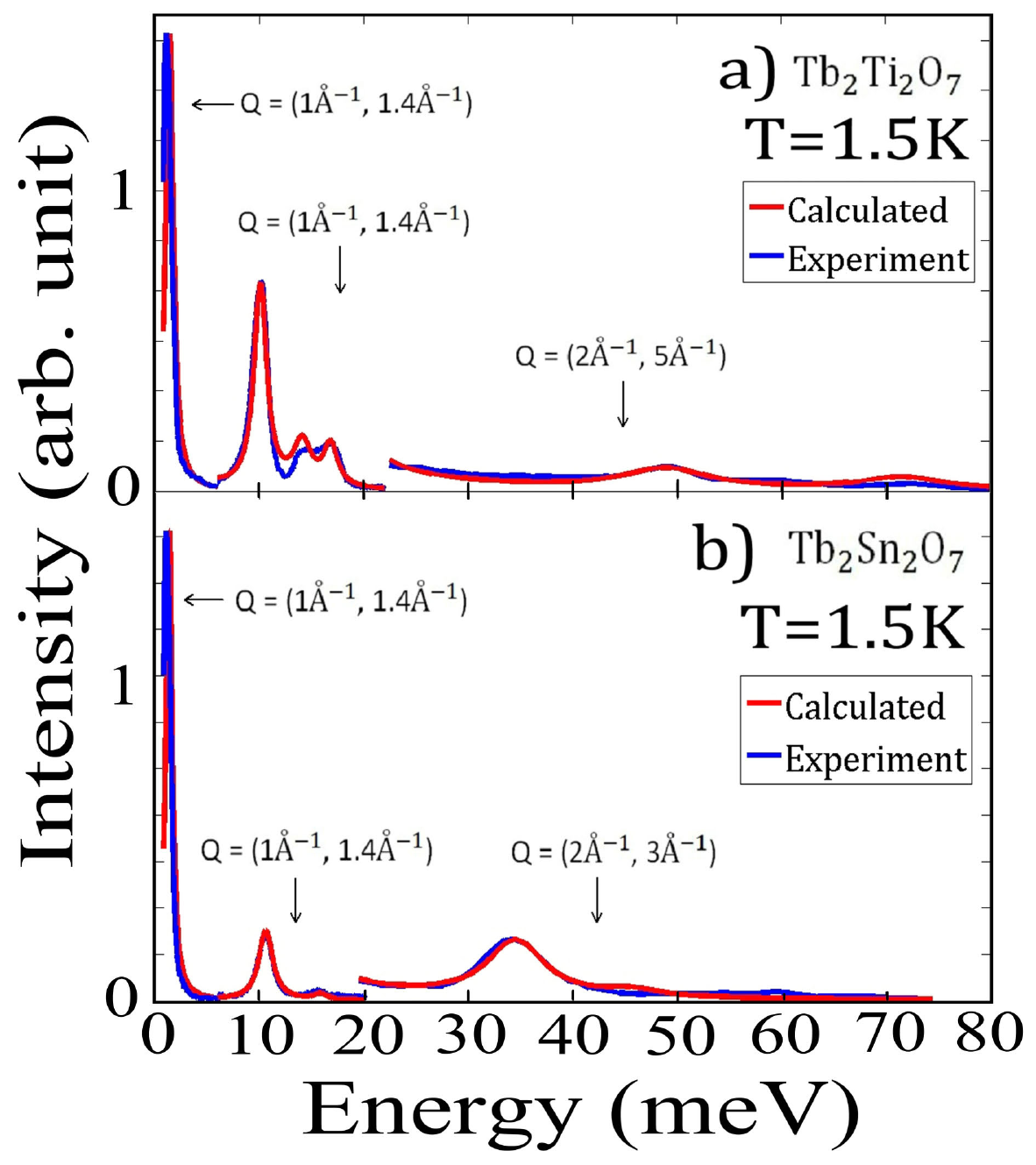}
\caption{(Color online) Cuts (blue) of the inelastic neutron scattering data shown in Fig. 2 are compared with theoretical calculations (red) of the magnetic neutron scattering cross section for crystal field excitations from (a) Tb$_2$Ti$_2$O$_7$ and (b) Tb$_2$Sn$_2$O$_7$, at $T=1.5$~K. Spectra over the three different energy ranges shown in Fig. 3 were taken using the three different $E_i$ shown in Fig. 2, and used an integration in $\vert {\bf Q} \vert$ as indicated in the figure. }
\end{figure}

In order to quantitatively analyze the neutron scattering results shown in Fig. 2 in terms of appropriate crystal field excitations discussed in  Sec. III, we need to extract line scans of intensity versus energy for data sets corresponding to each of the three $E_i$'s and for each of Tb$_2$Ti$_2$O$_7$ and Tb$_2$Sn$_2$O$_7$. Doing so requires an integration of the data in Fig. 2 in $\vert {\bf Q} \vert $. Due to the kinematic constraints of the neutron scattering, wherein it is impossible to satisfy energy and momentum conservation at small $\vert {\bf Q} \vert$ and high energy, such an integration in $\vert {\bf Q} \vert$ depends on the energy range of the crystal field excitation of interest, with more energetic crystal field excitations requiring access to higher $\vert {\bf Q} \vert$.  These results are shown in Fig. 3(a) for Tb$_2$Ti$_2$O$_7$ and in Fig. 3(b) for Tb$_2$Sn$_2$O$_7$. For both materials, we employ an integration over 1 \AA$^{-1}<$ $|{\bf Q}| <$ 1.4 \AA$^{-1}$ for both $E_i=11$ and 45 meV.  The highest energy crystal field excitations extend to higher energies in Tb$_2$Ti$_2$O$_7$ compared with Tb$_2$Sn$_2$O$_7$ and consequently for the $E_i=120$ meV data sets, we employ an integration of 2 \AA$^{-1}<|{\bf Q}| < 3$ \AA$^{-1}$ for Tb$_2$Sn$_2$O$_7$ and 2 \AA$^{-1}< |{\bf Q}|<$ 5 \AA$^{-1}$ for Tb$_2$Ti$_2$O$_7$.

\begin{figure}[h]
\includegraphics[width=\columnwidth]{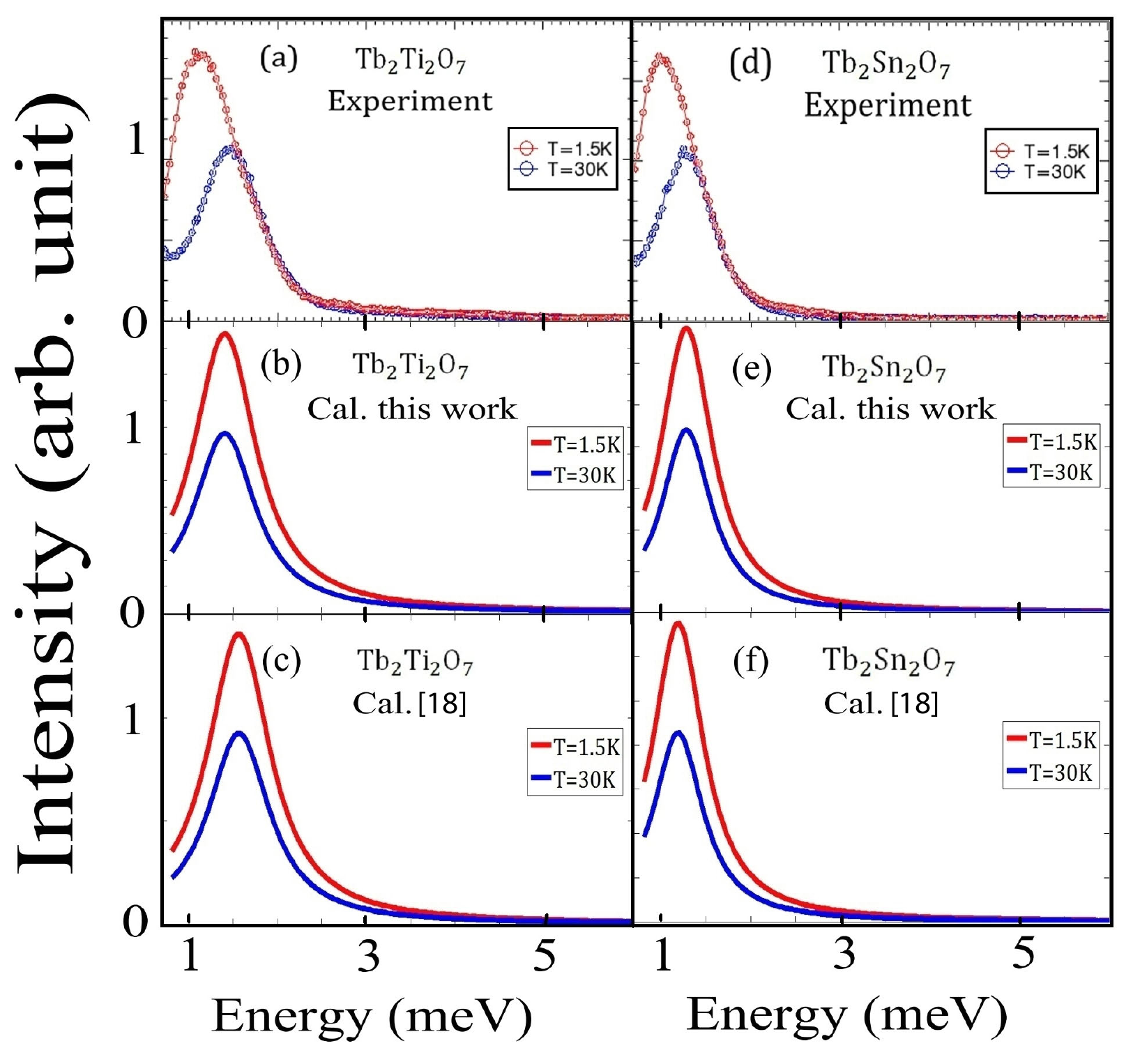}
\caption{(Color online) Comparison between the low-energy inelastic scattering measurements ($<6$ meV) and calculations for the magnetic neutron scattering between crystal field states for (a)-(c) Tb$_2$Ti$_2$O$_7$ and (d)-(f) Tb$_2$Sn$_2$O$_7$. This shows the measured and calculated spectra at both $T=1.5$~K and 30 K. The calculations shown in (b) and (e) use our newly determined crystal field parameters shown Table $\textrm{II}$. For comparison, (c) and (f) show the equivalent calculation using the crystal field parameters determined previously by Mirebeau \textit{et al.} \cite{Mirebeau2007}  As can be seen, the temperature dependence of the low-energy spectra does not distinguish between these two model calculations.}
\end{figure}

\begin{figure}[h]
\includegraphics[width=\columnwidth]{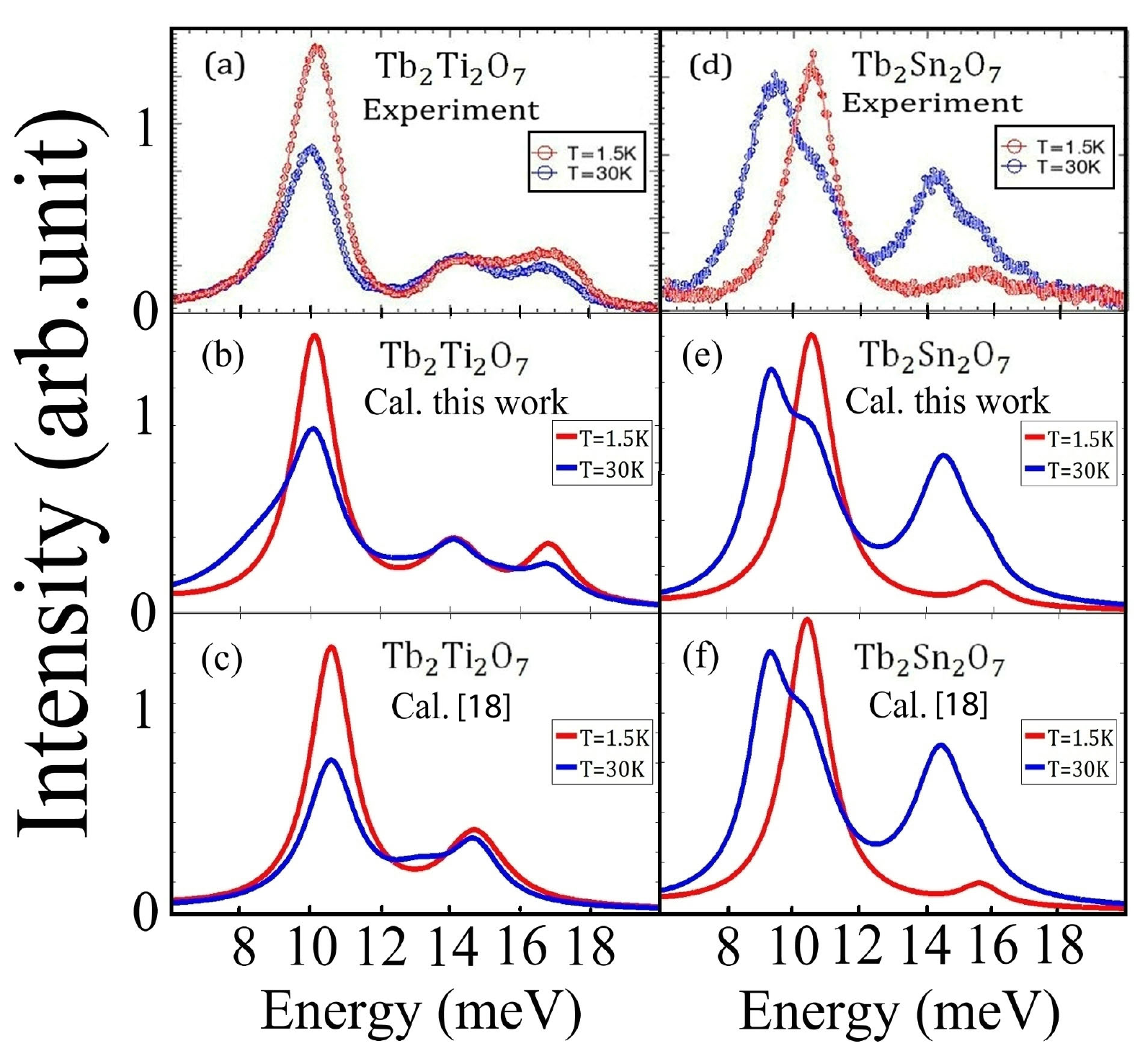}
\caption{(Color online) Comparison between inelastic scattering measurements at moderate energies (6 meV$<$ $E$ $<$ 20 meV) and calculations for the magnetic neutron scattering between crystal field states for (a)-(c) Tb$_2$Ti$_2$O$_7$ and (d)-(f) Tb$_2$Sn$_2$O$_7$. This shows the measured and calculated spectra at both $T=1.5$ K and 30 K. The calculations shown in (b) and (e) use our newly determined crystal field parameters shown in Table $\textrm{II}$.  For comparison, (c) and (f) show the equivalent calculation using the crystal field parameters determined previously by Mirebeau \textit{et al.}\cite{Mirebeau2007} At low temperatures, $T=1.5$~K, this energy regime shows transitions from both the ground-state doublet to crystal field states beyond the first excited state doublet.  At $T=30$~K, it shows transitions from both the ground-state doublet and first excited state doublet to crystal field states beyond the first excited state doublet. For that reason, and in contrast to data shown in Fig. 4, this inelastic data is very sensitive to the specific details of the spectral decomposition (eigenfunctions) of both the ground-state and first excited state doublets.}
\end{figure}

Figures 4(a), 4(d), 5(a), and 5(d) show the same data as was shown in Fig. 3 at $T=1.5$~K for (a) Tb$_2$Ti$_2$O$_7$ and (d) Tb$_2$Sn$_2$O$_7$, as well as the corresponding data at $T=30$~K for different energies. Figs. 4(a) and 4(d) show $E_i=11$ meV data sets from 1 to 6 meV, while Figs. 5(a) and 5(d) show the $E_i=45$ meV data sets from 7 to 20 meV. Taken together, Figs. 3(a), 4(a), and 5(a) show, at $T=1.5$~K, a single transition out of the ground state at $\sim1$ meV, one transition near 10 meV, and two close together near 14 and 16.5 meV, a transition near 49 meV and a final very weak transition near 70 meV for Tb$_2$Ti$_2$O$_7$.  For Tb$_2$Sn$_2$O$_7$ at $T=1.5$~K, we see a transition out of the ground state near 1 meV, another transition near 10.5 meV, a single transition near 15.5 meV, and a final observable transition at $\sim34$~meV. For both Tb$_2$Ti$_2$O$_7$ and Tb$_2$Sn$_2$O$_7$, the general trend is that the crystal field transitions are weaker at higher energies, and thus the identification of the transitions from the ground state to the excited crystal field states is more robust at lower energies. Nonetheless, the intensity of the inelastic peaks as a function of energy can be modeled by calculations of the magnetic neutron scattering cross section appropriate to transitions between Tb$^{3+}$ crystal field states, as described in Sect. III.  

These calculations have been carried out. In all cases, we set the energy widths associated with all transitions, $\Gamma _{\lambda \lambda '}$, to be determined by our finite energy resolution of $\sim3\%$ of $E_i$, so full widths at half maximum of $\sim0.3$~meV for $E_i$ = 11 meV, 1.4 meV for $E_i$ = 45 meV and 3.6 meV for $E_i$ = 120 meV. This calculation will not reproduce the dispersion known to be relevant to the lowest $\sim1.5$~meV transitions at $T=1.5$~K, in both materials, as shown in Figs. 4(a) and 4(d). For that reason, we fitted the integrated intensity of this lowest transition by artifically raising its energy to its 1.5 meV value observed at $T=30$~K. 

In the case of Tb$_2$Sn$_2$O$_7$, we used the crystal field parameters previously determined for Tb$_2$Sn$_2$O$_7$ by Mirebeau and co-workers as starting points to fit our spectra. \cite{Mirebeau2007} In the case of Tb$_2$Ti$_2$O$_7$, we also used the Mirebeau \textit{et al.} \cite{Mirebeau2007} parameters determined for Tb$_2$Ti$_2$O$_7$, but minimized both the sum of the least squares of the difference between the measured and calculated neutron intensity, as well as the sum of the least squares of the difference between the experimental and calculated crystal field energies (subject to the aforementioned caveat regarding the dispersion of the lowest lying crystal field excitation). In the case of Tb$_2$Ti$_2$O$_7$, this was required in order to place {\it two excitations} from the ground-state doublet to excited states in the 12-17 meV energy range at $T=1.5$~K, as observed in the data reported here [see Fig. 5 (a)], and also in earlier measurements on Tb$_2$Ti$_2$O$_7$.  

The crystal field calculation using the Tb$_2$Ti$_2$O$_7$ parameters determined by Mirebeau \textit{et al.} \cite{Mirebeau2007} produces only a single crystal field level in this same energy regime at $T=1.5$~K. For this reason, and as shown explicitly in Fig. 5(c), the Tb$_2$Ti$_2$O$_7$ calculated parameters determined from Mirebeau \textit{et al.} \cite{Mirebeau2007} do not describe our measurements well in this 7-20 meV regime. This is in contrast to our Tb$_2$Sn$_2$O$_7$ neutron data, where the Mirebeau \textit{et al.} parameters do provide a good description of our data,  as can be seen by comparing Figs. 5(d)-5(f).

The results of this fitting procedure are shown as the red lines in Figs. 3(a) and 3(b) for Tb$_2$Ti$_2$O$_7$ and Tb$_2$Sn$_2$O$_7$, respectively, and in Figs. 4(b) and 5(b) for Tb$_2$Ti$_2$O$_7$ at $T=1.5$ and 30 K, as well as Figs. 4(e) and 5(e) for Tb$_2$Sn$_2$O$_7$ at $T=1.5$ K and 30 K. Clearly, the description of the crystal field excitation energies and intensities is good over the full energy range studied for both Tb$_2$Ti$_2$O$_7$ and Tb$_2$Sn$_2$O$_7$.  We also repeated this same fitting protocol for both materials using a variable energy width, $\Gamma _{\lambda \lambda '}$, for each transition.  While this obviously produced better agreement between the measured and calculated neutron spectra than fitting with $\Gamma _{\lambda \lambda '}$ fixed at values approximating the finite energy resolution of the spectrometer, it did not affect the eigenvalues or eigenfunctions extracted from these fits beyond a $\pm2\%$ level.

It is clear from Figs. 2(a) and 2(d) that the low-energy crystal field eigenvalues are very similar in Tb$_2$Ti$_2$O$_7$ and Tb$_2$Sn$_2$O$_7$ at both $T=1.5$ and $T=30$ K. The mean energy and the weak dispersion characterizing the transition from the ground state to first excited state are both very similar in the two materials. It is at higher energies, in particular between 6 and 20 meV, as shown in Fig. 5(a) and 5(d), where significant qualitative differences appear. Specifically at $T=1.5$~K, Tb$_2$Ti$_2$O$_7$ shows three transitions from the ground state to excited crystal field states in this range, while Tb$_2$Sn$_2$O$_7$ shows only two. This difference, clear in both the $T=1.5$~K data sets of Fig. 5(a) and 5(d), and in the very different temperature dependence which occurs on populating the first set of excited states, as occurs for either material at $T=30$~K, was not fully appreciated in earlier triple axis neutron spectroscopic studies of the Tb$^{3+}$ levels in Tb$_2$Ti$_2$O$_7$ \cite{Mirebeau2007,Gingras2000} and Tb$_2$Sn$_2$O$_7$. \cite{Mirebeau2007} The first such work, by Gingras \textit{et al.} \cite{Gingras2000} on Tb$_2$Ti$_2$O$_7$ was performed at $T=12$~K and consequently the ``extra" peak near 14 meV was misinterpreted as arising out of the first excited state, rather than out of the ground state. The later work by Mirebeau \textit{et al.} \cite{Mirebeau2007} on Tb$_2$Ti$_2$O$_7$ did show a discrepancy between triple axis neutron measurements and the corresponding crystal field calculation. However, this discrepancy was not refined such that the three excitations at $T=1.5$~K, from the ground-state doublet of Tb$_2$Ti$_2$O$_7$ to excited crystal field states in the 7-20 meV range, could be accounted for.

Our fits to our new Tb$_2$Ti$_2$O$_7$ data in this range, shown in Figs. 5(a) and 5(b), are an excellent description of this data at $T=1.5$~K. At $T=30$~K, the description is also good, although a weak excitation arising from a transition from the excited state doublet to the singlet, $\sim10$ meV above the ground state is not as evident in the data [Fig. 5(a)] as the calculation [Fig. 5(b)] predicts. There is a small shift towards lower energies in the peak near 10 meV at 30 K compared with 1.5 K, and the line shape is asymmetric with enhanced scattering on the lower-energy side, consistent with an additional peak on the low-energy shoulder of the $\sim10$ meV peak at $T=30$~K. Again, the observed effect is less pronounced than the calculation. This could arise because this excitation out of the excited state doublet has either dispersion or a finite lifetime, and therefore has a larger energy width than the one that we have ascribed to it on the basis of energy resolution alone. Indeed, the aforementioned fits in which the $\Gamma _{\lambda \lambda '}$ is a fit parameter account for this effect well.

The six crystal field parameters, $D^m_n$ in Eq. (1), obtained from this fitting procedure for Tb$_2$Ti$_2$O$_7$ and Tb$_2$Sn$_2$O$_7$ are listed in Table II [in units of wave number cm$^{-1}$, 1 K $= 0.0862$ meV$ = 0.695$ cm$^{-1}$]. This set of parameters describes very well the transitions out of the ground state (at $T=1.5$~K) shown in Figs. 3-5, as well as out of the equilibrium population of ground and first excited states appropriate to $T=30$~K. Given that the first excited state is at an energy of $\sim1.5$ meV when $T=30$~K, and the next highest-energy crystal field state is at $\sim10$ meV $\sim110$ K, both the ground state and first excited states in Tb$_2$Ti$_2$O$_7$ and Tb$_2$Sn$_2$O$_7$ are approximately equally populated at $T=30$~K. The calculation of the magnetic neutron scattering in the low-energy regime, with transitions between the ground and first excited states only, is shown in Fig. 4. The intensity of this transition drops by a factor of $\sim2$ between $T=1.5$~K and $T=30$~K for any model that possesses the same degeneracy for the ground and first excited states. This is the case for both our crystal field parameters shown in Table II and those obtained earlier by Mirebeau and co-workers.\cite{Mirebeau2007} Thus, the temperature dependence of the low-energy crystal field scattering is not a sensitive probe of the nature of the Tb$^{3+}$ crystal field parameters in either material. However, the temperature dependence of the crystal field scattering at higher energies, as shown between 6 and 20 meV in Fig. 5, such that one can also probe transitions from the first excited state, which is populated at 30 K, is much richer. Here, one is very sensitive to the {\it concurrent} details of the ground-state and first excited state eigenfunctions, as well as to the details of the eigenfunctions relevant to the high-energy crystal field states to which the neutron can make transitions to.

\begin{table}[t!]
\newcommand{\tabincell}[2]{\begin{tabular}{@{}#1@{}}#2\end{tabular}}
\caption{Crystal field parameters ($D_n^m$) appropriate to Tb$^{3+}$ in Tb$_2$Ti$_2$O$_7$ and Tb$_2$Sn$_2$O$_7$ as derived from fitting our inelastic neutron scattering data at both $T=1.5$ K and 30 K with calculations of the magnetic neutron scattering cross section due to transitions between crystal field states. Note that 1 K $=0.0862$ meV$=$ 0.695 cm$^{-1}$.}
\begin{tabular*}{8.7cm}{{@{\extracolsep{\fill}}|c|c|c|c|c|c|c|}}
\hline \hline
 &\tabincell{c}{$D_2^0$\\(K)}&\tabincell{c}{$D_4^0$\\(K)}&\tabincell{c}{$D_4^3$\\(K)}&\tabincell{c}{$D_6^0$\\(K)}&\tabincell{c}{$D_6^3$\\(K)}&\tabincell{c}{$D_6^6$\\(K)}\\
\hline 
Tb$_2$Ti$_2$O$_7$&833.12&389.39&$5562.84$&123.98&$5182.89$&$8807.51$\\
\hline
Tb$_2$Sn$_2$O$_7$&225.37&331.9&249.59&-29.17&-649.79&1199.91\\
\hline\hline
\end{tabular*}
\end{table}

\section{DISCUSSION}

\subsection{Crystal structure and crystal field parameters}

The resulting energy eigenvalues and eigenfunctions that arise from fitting our inelastic neutron scattering data to Tb$_2$Ti$_2$O$_7$ and Tb$_2$Sn$_2$O$_7$ are given in Table III and the spectra of energy eigenvalues are illustrated in Fig. 6.  As was clear from our earlier discussion, and consistent with earlier measurements, the lowest-energy sector consists of a ground-state doublet and first excited state doublet separated by an energy $\Delta_1\sim1.5$ meV in both materials. However, the eigenfunctions making these doublets are quite different. In the case of Tb$_2$Sn$_2$O$_7$, our results are fully consistent with early triple axis work, \cite{Mirebeau2007} wherein the ground state ($g$) and first excited state ($1$) doublets are described by

\begin{equation}
\begin{split}
\left|\psi_g^{\rm Sn}\right\rangle=&0.925\left|\mp5\right\rangle\pm0.292\left|\pm4\right\rangle\pm0.242\left|\mp2\right\rangle\\
&+0.017\left|\pm1\right\rangle
\end{split}
\end{equation}

and

\begin{equation}
\begin{split}
\left|\psi_1^{\rm Sn} \right\rangle=&0.942\left|\mp4\right\rangle\pm0.300\left|\pm5\right\rangle\pm0.149\left|\mp1\right\rangle\\
&+0.020\left|\pm2\right\rangle
\end{split}
\end{equation}

For Tb$_2$Ti$_2$O$_7$, however, our new eigenfunctions are very different from those previously reported, \cite{Mirebeau2007,Gingras2000} and are given by

\begin{equation}
\begin{split}
\left|\psi_g^{\rm Ti} \right\rangle=&0.810\left|\mp5\right\rangle\pm0.472\left|\pm4\right\rangle\pm0.338\left|\mp2\right\rangle\\
&+0.078\left|\pm1\right\rangle
\end{split}
\end{equation}

and

\begin{equation}
\begin{split}
\left|\psi_1^{\rm Ti} \right\rangle=&0.799\left|\mp 4\right\rangle\pm0.507\left|\pm5\right\rangle\pm0.279\left|\mp1\right\rangle\\
&+0.163\left|\pm2\right\rangle
\end{split}
\end{equation}

\begin{table*}[t!]
\newcommand{\tabincell}[2]{\begin{tabular}{@{}#1@{}}#2\end{tabular}}
\caption{The crystal field eigenvalues and eigenstates determined for Tb$^{3+}$ in Tb$_2$Ti$_2$O$_7$ (top) and Tb$_2$Sn$_2$O$_7$ (bottom) are shown. The first column displays all of the crystal field energy eigenvalues, while the corresponding eigenfunction appropriate to each eigenvalue is given in each such row, in terms of the $J_z$ basis states for $J=6$.}
\begin{tabular*}{17cm}{{@{\extracolsep{\fill}}|l|ccccccccccccc|}}
\hline \hline
\tabincell{c}{Tb$_2$Ti$_2$O$_7$\\E(meV)}&$\left|-6\right\rangle$&$\left|-5\right\rangle$&$\left|-4\right\rangle$&$\left|-3\right\rangle$&$\left|-2\right\rangle$&$\left|-1\right\rangle$&$\left|0\right\rangle$&$\left|1\right\rangle$&$\left|2\right\rangle$&$\left|3\right\rangle$&$\left|4\right\rangle$&$\left|5\right\rangle$&$\left|6\right\rangle$\\
\hline 
0&0&0.810&0&0&0.338&0&0&0.078&0&0&0.472&0&0\\
\hline
0&0&0&-0.472&0&0&0.078&0&0&-0.338&0&0&0.810&0\\
\hline
1.41&0&-0.507&0&0&0.163&0&0&-0.279&0&0&0.799&0&0\\
\hline 
1.41&0&0&0.799&0&0&0.279&0&0&0.163&0&0&0.507&0\\
\hline 
10.12&0.055&0&0&0.705&0&0&0&0&0&0.705&0&0&-0.055\\
\hline 
14.13&0.689&0&0&0.084&0&0&0.193&0&0&-0.084&0&0&0.689\\
\hline 
16.85&-0.705&0&0&0.055&0&0&0&0&0&0.055&0&0&0.705\\
\hline 
48.96&0.136&0&0&-0.623&0&0&-0.432&0&0&0.623&0&0&0.136\\
\hline 
71.68&0&0&-0.217&0&0&-0.372&0&0&0.861&0&0&0.269&0\\
\hline
71.68&0&-0.269&0&0&0.861&0&0&0.372&0&0&-0.217&0&0\\
\hline 
100.99&0&0&-0.302&0&0&0.882&0&0&0.342&0&0&-0.118&0\\
\hline
100.99&0&-0.118&0&0&-0.342&0&0&0.882&0&0&0.302&0&0\\
\hline 
113.20&-0.084&0&0&-0.324&0&0&0.881&0&0&0.324&0&0&-0.084\\
\hline\hline
\tabincell{c}{Tb$_2$Sn$_2$O$_7$\\E(meV)}&$\left|-6\right\rangle$&$\left|-5\right\rangle$&$\left|-4\right\rangle$&$\left|-3\right\rangle$&$\left|-2\right\rangle$&$\left|-1\right\rangle$&$\left|0\right\rangle$&$\left|1\right\rangle$&$\left|2\right\rangle$&$\left|3\right\rangle$&$\left|4\right\rangle$&$\left|5\right\rangle$&$\left|6\right\rangle$\\
\hline
0&0&0.925&0&0&0.242&0&0&0.017&0&0&0.292&0&0\\
\hline 
0&0&0&-0.292&0&0&0.017&0&0&-0.242&0&0&0.925&0\\
\hline 
1.28&0&0&0.942&0&0&0.149&0&0&0.020&0&0&0.300&0\\
\hline
1.28&0&-0.300&0&0&0.020&0&0&-0.149&0&0&0.942&0&0\\
\hline 
10.55&0.207&0&0&0.676&0&0&0&0&0&0.676&0&0&-0.207\\
\hline 
15.80&-0.273&0&0&-0.641&0&0&-0.170&0&0&0.641&0&0&-0.273\\
\hline 
34.00&-0.676&0&0&0.207&0&0&0&0&0&0.207&0&0&0.676\\
\hline
34.21&0&-0.229&0&0&0.964&0&0&0.113&0&0&-0.075&0&0\\
\hline 
34.21&0&0&-0.076&0&0&-0.031&0&0&0.964&0&0&0.230&0\\
\hline 
34.81&0.652&0&0&-0.274&0&0&-0.031&0&0&0.274&0&0&0.652\\
\hline 
45.56&0&0&-0.147&0&0&0.982&0&0&0.112&0&0&-0.035&0\\
\hline
45.56&0&-0.035&0&0&-0.112&0&0&0.982&0&0&0.147&0&0\\
\hline 
49.81&-0.027&0&0&-0.119&0&0&0.985&0&0&0.119&0&0&-0.027\\
\hline\hline
\end{tabular*}
\end{table*}

\begin{figure}[h]
\includegraphics[width=\columnwidth]{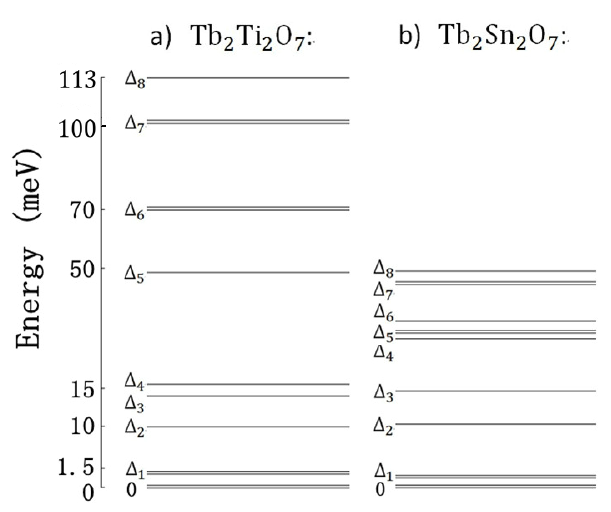}
\caption{Crystal field energy scheme for all the $2J+1=13$ levels for Tb$^{3+}$ in (a) Tb$_2$Ti$_2$O$_7$ and (b) Tb$_2$Sn$_2$O$_7$, as derived from our calculations of the magnetic neutron scattering cross section from transitions between crystal field states (not to scale). These calculations employed the best-fit crystal field parameter sets determined for Tb$_2$Ti$_2$O$_7$ and Tb$_2$Sn$_2$O$_7$ shown in Table $\textrm{II}$.}
\end{figure}

For reference, we have calculated the magnetic neutron scattering spectra using the crystal field parameters obtained earlier by Mirebeau \textit{et al.}, \cite{Mirebeau2007} and these results are compared with
our new experimental data in Figs. 4(c), 4(f), 5(c) and 5(f). It is only for Tb$_2$Ti$_2$O$_7$ and only for the crystal field levels in the 6 to 20 meV range, shown in Fig. 5(c), that the deficiency of the earlier estimate for the crystal field parameters is evident. We note that earlier still triple axis work by Gingras \textit{et al.}, \cite{Gingras2000} which had reached much the same conclusions as the later work of Mirebeau \textit{et al.}, \cite{Mirebeau2007} did not produce estimates for the $D^m_n$ crystal field parameters, and therefore we could not compare these earlier calculations to our new experiment.

We note that Fig. 6, which illustrates the dynamic range of the splitting of the $2J+1=13$ crystal field levels in Tb$_2$Ti$_2$O$_7$ and Tb$_2$Sn$_2$O$_7$, makes it clear that the overall scale or bandwidth for the crystal field splitting is a factor of $\sim2$ greater in Tb$_2$Ti$_2$O$_7$ compared with Tb$_2$Sn$_2$O$_7$. There are two free parameters within the $Fd\bar{3}m$ space group that Tb$_2$Ti$_2$O$_7$ and Tb$_2$Sn$_2$O$_7$ share: the lattice parameter of the conventional unit cell $a$ and the oxygen position parameter $x$, which determines the positions of eight oxygen ions in the immediate vicinity of the Tb$^{3+}$ ion.  It is the cubic lattice parameter, $a$ which, at room temperature, increases from Tb$_2$Ti$_2$O$_7$ ($a=10.149$~\AA) to Tb$_2$Sn$_2$O$_7$ ($a=10.426$~\AA) that is largely responsible for the fact that the overall bandwidth of crystal field splittings is smaller in Tb$_2$Sn$_2$O$_7$ compared with Tb$_2$Ti$_2$O$_7$. Such an effect is expected; nonetheless, its scale is remarkably large.

Our newly determined eigenfunctions for the ground-state and first excited crystal field state doublets in Tb$_2$Ti$_2$O$_7$ are more similar to those appropriate to Tb$_2$Sn$_2$O$_7$ than previously thought.  As shown in Eqs. (6) and (7) and Table III, the ground-state doublet in Tb$_2$Sn$_2$O$_7$ remains primarily  $m_J=\left|\pm 5\right\rangle$, while the first excited state doublet remains primarily  $m_J=\left|\pm 4\right\rangle$. For Tb$_2$Ti$_2$O$_7$, the ground-state and first excited state doublets are now determined to be a more equitable mixture of the symmetry allowed $m_J$ basis states, with the largest component being  $m_J=\left|\pm 5\right\rangle$ for the ground-state and  $m_J=\left|\pm 4\right\rangle$ for the first excited state doublets. These new eigenfunctions also allow us to calculate the ground-state magnetic moment for both Tb$_2$Ti$_2$O$_7$ and Tb$_2$Sn$_2$O$_7$, which are 3.92(8)$\mu_{\rm B}$ and 6.1(1)$\mu_{\rm B}$, respectively.

While the effective exchange interactions between Tb$^{3+}$ rare-earth moments are expected to be anisotropic in nature, as found for both  Yb$_2$Ti$_2$O$_7$ \cite{Ross2011} and Er$_2$Ti$_2$O$_7$,\cite{Savary2012} and recently discussed for the case of Tb$_2$Ti$_2$O$_7$, \cite{Curnoe2013} the differences between the microscopic Hamiltonians in Tb$_2$Ti$_2$O$_7$ and Tb$_2$Sn$_2$O$_7$ are likely to be subtle. Tb$_2$Ti$_2$O$_7$ has the more antiferromagnetic $\Theta_{\rm CW}$ and its ground-state moment should be slightly smaller than that of Tb$_2$Sn$_2$O$_7$, a consequence of the significant admixing of small $m_J$ basis states into the eigenfunction of the ground-state doublet. The smaller magnetic moment translates into weaker dipolar interactions. As the near-neighbor part of the dipolar interactions on the pyrochlore lattice is effectively ferromagnetic in nature, both of these effects will tend to make Tb$_2$Sn$_2$O$_7$ closer to the expectations of a dipolar spin ice ground state.\cite{Hertog2000}  Recent neutron experiments are now showing a disordered antiferromagnetically correlated spin ice state in Tb$_2$Ti$_2$O$_7$ at very low temperatures in zero field.\cite{Fritsch2013} Therefore both materials exhibit a variant of the spin ice ground state at sufficiently low temperatures, which is consistent with the similarities between their structures, low-lying crystal field states, and their interactions as quantified through their $\Theta_{\rm CW}$. 

\section{CONCLUSIONS}

We have performed new time-of-flight inelastic neutron scattering spectroscopy experiments to characterize the crystalline electric field states of Tb$^{3+}$ in Tb$_2$Ti$_2$O$_7$ and Tb$_2$Sn$_2$O$_7$. These measurements probe transitions out of the ground-state doublet alone at $T=1.5$~K, and out of the equilibrium distribution of the ground-state and first excited state doublets at $T=30$~K. We use these magnetic excitation spectra to produce a robust determination of the eigenvalues and eigenfunctions of the $2J+1=13$ crystal field states for the $J=6$ Tb$^{3+}$ magnetic moments in these materials. Our results for Tb$_2$Sn$_2$O$_7$ are consistent with a previous determination of these eigenfunctions and eigenvalues, for which the ground-state doublet is principally comprised of the  $m_J=\left|\pm 5\right\rangle$ basis states and the first excited state is principally comprised of the  $m_J=\left|\pm 4\right\rangle$ basis states. Our results for Tb$_2$Ti$_2$O$_7$ are not consistent with previous estimates, mainly due to a previously misidentified transition from the ground-state doublet near 14 meV. Our new determination of the ground-state and first excited state
eigenfunctions in Tb$_2$Ti$_2$O$_7$ show them to be made up of a distribution of the symmetry allowed $m_J$ basis states, where the largest contribution to the ground-state doublet comes from  $m_J=\left|\pm 5\right\rangle$, and the largest contribution to the first excited state doublet comes from  $m_J=\left|\pm 4\right\rangle$. This is in contrast with previous works,\cite{Mirebeau2007,Gingras2000} which had found the ground and excited doublets to be, respectively, predominantly made of $m_J = \vert \pm 4\rangle$ and
$m_J = \vert \pm 5 \rangle$. The detailed eigenfunctions and eigenvalues for Tb$_2$Ti$_2$O$_7$ and  Tb$_2$Sn$_2$O$_7$ are listed in Table III and illustrated in Fig. 6.

These results show the bandwidth of the energy eigenvalues to be a remarkable factor of $\sim2$ greater in Tb$_2$Ti$_2$O$_7$ compared with Tb$_2$Sn$_2$O$_7$. This is qualitatively consistent with the fact that the cubic pyrochlore lattice is expanded in Tb$_2$Sn$_2$O$_7$ compared with Tb$_2$Ti$_2$O$_7$.

Finally, we note that the precise nature of the low-lying crystal field levels is central to both theoretical proposals\cite{Molavian2007,Bonville2011} which have been put forward to explain the lack of magnetic order in Tb$_2$Ti$_2$O$_7$ to temperatures much less than 1 K. Our results will help contribute to the construction of a more precise theoretical model of Tb$_2$Ti$_2$O$_7$, which is vital to the potential resolution of the debate of the nature of its perplexing ground state.

\section*{Acknowledgments}
Research at the SNS at Oak Ridge National Laboratory was sponsored by the Scientific User Facilities Division, Office of Basic Energy Sciences, US Department of Energy. Work at McMaster University was supported by NSERC of Canada. Work at Princeton University was supported by grant DE-FG02-08ER46544. MJPG acknowledges the Canada Research Chair program, the Canada Council for the Arts for a Killam Research Fellowship and the Perimeter Institute for Theoretical Physics. Research at the Perimeter Institute is supported by the Government of Canada through Industry Canada and by the Province of Ontario through the Ministry of Economic Development \& Innovation. The neutron scattering data were reduced using Mantid \cite{Mantid} and analyzed using the Dave software package.\cite{DAVE}


\begin{thebibliography}{69}%
\makeatletter
\providecommand \@ifxundefined [1]{%
 \@ifx{#1\undefined}
}%
\providecommand \@ifnum [1]{%
 \ifnum #1\expandafter \@firstoftwo
 \else \expandafter \@secondoftwo
 \fi
}%
\providecommand \@ifx [1]{%
 \ifx #1\expandafter \@firstoftwo
 \else \expandafter \@secondoftwo
 \fi
}%
\providecommand \natexlab [1]{#1}%
\providecommand \enquote  [1]{``#1''}%
\providecommand \bibnamefont  [1]{#1}%
\providecommand \bibfnamefont [1]{#1}%
\providecommand \citenamefont [1]{#1}%
\providecommand \href@noop [0]{\@secondoftwo}%
\providecommand \href [0]{\begingroup \@sanitize@url \@href}%
\providecommand \@href[1]{\@@startlink{#1}\@@href}%
\providecommand \@@href[1]{\endgroup#1\@@endlink}%
\providecommand \@sanitize@url [0]{\catcode `\\12\catcode `\$12\catcode
  `\&12\catcode `\#12\catcode `\^12\catcode `\_12\catcode `\%12\relax}%
\providecommand \@@startlink[1]{}%
\providecommand \@@endlink[0]{}%
\providecommand \url  [0]{\begingroup\@sanitize@url \@url }%
\providecommand \@url [1]{\endgroup\@href {#1}{\urlprefix }}%
\providecommand \urlprefix  [0]{URL }%
\providecommand \Eprint [0]{\href }%
\providecommand \doibase [0]{http://dx.doi.org/}%
\providecommand \selectlanguage [0]{\@gobble}%
\providecommand \bibinfo  [0]{\@secondoftwo}%
\providecommand \bibfield  [0]{\@secondoftwo}%
\providecommand \translation [1]{[#1]}%
\providecommand \BibitemOpen [0]{}%
\providecommand \bibitemStop [0]{}%
\providecommand \bibitemNoStop [0]{.\EOS\space}%
\providecommand \EOS [0]{\spacefactor3000\relax}%
\providecommand \BibitemShut  [1]{\csname bibitem#1\endcsname}%
\let\auto@bib@innerbib\@empty
\bibitem [{\citenamefont {Lacroix}\ \emph {et~al.}(2011)\citenamefont
  {Lacroix}, \citenamefont {Mendels},\ and\ \citenamefont
  {Mila}}]{Lacroix2011}%
  \BibitemOpen
  \bibfield  {author} {\bibinfo {author} {\bibfnamefont {C.}~\bibnamefont
  {Lacroix}}, \bibinfo {author} {\bibfnamefont {P.}~\bibnamefont {Mendels}}, \
  and\ \bibinfo {author} {\bibfnamefont {F.}~\bibnamefont {Mila}},\ }\enquote
  {\bibinfo {title} {Introduction to frustrated magnetism},}\ \ (\bibinfo
  {publisher} {Springer Series in Solid-State Sciences},\ \bibinfo {address}
  {Heidelberg},\ \bibinfo {year} {2011})\BibitemShut {NoStop}%
\bibitem [{\citenamefont {Kadowaki}\ \emph {et~al.}(1987)\citenamefont
  {Kadowaki}, \citenamefont {Ubukoshi}, \citenamefont {Hirakawa}, \citenamefont
  {Mart\'{i}nez},\ and\ \citenamefont {Shirane}}]{Kadowaki1987}%
  \BibitemOpen
  \bibfield  {author} {\bibinfo {author} {\bibfnamefont {H.}~\bibnamefont
  {Kadowaki}}, \bibinfo {author} {\bibfnamefont {K.}~\bibnamefont {Ubukoshi}},
  \bibinfo {author} {\bibfnamefont {K.}~\bibnamefont {Hirakawa}}, \bibinfo
  {author} {\bibfnamefont {J.~L.}\ \bibnamefont {Mart\'{i}nez}}, \ and\
  \bibinfo {author} {\bibfnamefont {G.}~\bibnamefont {Shirane}},\ }\href
  {\doibase 10.1143/JPSJ.56.4027} {\bibfield  {journal} {\bibinfo  {journal}
  {J. Phys. Soc. Jpn.}\ }\textbf {\bibinfo {volume} {56}},\ \bibinfo {pages}
  {4027} (\bibinfo {year} {1987})}\BibitemShut {NoStop}%
\bibitem [{\citenamefont {Collins}\ and\ \citenamefont
  {Petrenko}(1997)}]{Collins1997}%
  \BibitemOpen
  \bibfield  {author} {\bibinfo {author} {\bibfnamefont {M.~F.}\ \bibnamefont
  {Collins}}\ and\ \bibinfo {author} {\bibfnamefont {O.~A.}\ \bibnamefont
  {Petrenko}},\ }\href {\doibase 10.1139/p97-007} {\bibfield  {journal}
  {\bibinfo  {journal} {Can. J. Phys.}\ }\textbf {\bibinfo {volume} {75}},\
  \bibinfo {pages} {605} (\bibinfo {year} {1997})}\BibitemShut {NoStop}%
\bibitem [{\citenamefont {Inami}\ \emph {et~al.}(2000)\citenamefont {Inami},
  \citenamefont {Nishiyama}, \citenamefont {Maegawa},\ and\ \citenamefont
  {Oka}}]{Inami2000}%
  \BibitemOpen
  \bibfield  {author} {\bibinfo {author} {\bibfnamefont {T.}~\bibnamefont
  {Inami}}, \bibinfo {author} {\bibfnamefont {M.}~\bibnamefont {Nishiyama}},
  \bibinfo {author} {\bibfnamefont {S.}~\bibnamefont {Maegawa}}, \ and\
  \bibinfo {author} {\bibfnamefont {Y.}~\bibnamefont {Oka}},\ }\href {\doibase
  10.1103/PhysRevB.61.12181} {\bibfield  {journal} {\bibinfo  {journal} {Phys.
  Rev. B}\ }\textbf {\bibinfo {volume} {61}},\ \bibinfo {pages} {12181}
  (\bibinfo {year} {2000})}\BibitemShut {NoStop}%
\bibitem [{\citenamefont {Wills}\ \emph {et~al.}(2000)\citenamefont {Wills},
  \citenamefont {Harrison}, \citenamefont {Ritter},\ and\ \citenamefont
  {Smith}}]{Wills2000}%
  \BibitemOpen
  \bibfield  {author} {\bibinfo {author} {\bibfnamefont {A.~S.}\ \bibnamefont
  {Wills}}, \bibinfo {author} {\bibfnamefont {A.}~\bibnamefont {Harrison}},
  \bibinfo {author} {\bibfnamefont {C.}~\bibnamefont {Ritter}}, \ and\ \bibinfo
  {author} {\bibfnamefont {R.~I.}\ \bibnamefont {Smith}},\ }\href {\doibase
  10.1103/PhysRevB.61.6156} {\bibfield  {journal} {\bibinfo  {journal} {Phys.
  Rev. B}\ }\textbf {\bibinfo {volume} {61}},\ \bibinfo {pages} {6156}
  (\bibinfo {year} {2000})}\BibitemShut {NoStop}%
\bibitem [{\citenamefont {Han}\ \emph {et~al.}(2012)\citenamefont {Han},
  \citenamefont {Helton}, \citenamefont {Chu}, \citenamefont {Nocera},
  \citenamefont {Rodriguez-Rivera}, \citenamefont {Broholm},\ and\
  \citenamefont {Lee}}]{Han}%
  \BibitemOpen
  \bibfield  {author} {\bibinfo {author} {\bibfnamefont {T.-H.}\ \bibnamefont
  {Han}}, \bibinfo {author} {\bibfnamefont {J.~S.}\ \bibnamefont {Helton}},
  \bibinfo {author} {\bibfnamefont {S.}~\bibnamefont {Chu}}, \bibinfo {author}
  {\bibfnamefont {D.~G.}\ \bibnamefont {Nocera}}, \bibinfo {author}
  {\bibfnamefont {J.~A.}\ \bibnamefont {Rodriguez-Rivera}}, \bibinfo {author}
  {\bibfnamefont {C.}~\bibnamefont {Broholm}}, \ and\ \bibinfo {author}
  {\bibfnamefont {Y.~S.}\ \bibnamefont {Lee}},\ }\href {\doibase
  10.1038/nature11659} {\bibfield  {journal} {\bibinfo  {journal} {Nature (London)}\
  }\textbf {\bibinfo {volume} {492}},\ \bibinfo {pages} {406} (\bibinfo {year}
  {2012})}\BibitemShut {NoStop}%
\bibitem [{\citenamefont {Ramirez}\ \emph {et~al.}(1999)\citenamefont
  {Ramirez}, \citenamefont {Hayashi}, \citenamefont {Cava}, \citenamefont
  {Siddharthan},\ and\ \citenamefont {Shastry}}]{Ramirez1999}%
  \BibitemOpen
  \bibfield  {author} {\bibinfo {author} {\bibfnamefont {A.~P.}\ \bibnamefont
  {Ramirez}}, \bibinfo {author} {\bibfnamefont {A.}~\bibnamefont {Hayashi}},
  \bibinfo {author} {\bibfnamefont {R.~J.}\ \bibnamefont {Cava}}, \bibinfo
  {author} {\bibfnamefont {R.}~\bibnamefont {Siddharthan}}, \ and\ \bibinfo
  {author} {\bibfnamefont {B.~S.}\ \bibnamefont {Shastry}},\ }\href {\doibase
  10.1038/20619} {\bibfield  {journal} {\bibinfo  {journal} {Nature (London)}\ }\textbf
  {\bibinfo {volume} {399}},\ \bibinfo {pages} {333} (\bibinfo {year}
  {1999})}\BibitemShut {NoStop}%
\bibitem [{\citenamefont {Bramwell}\ and\ \citenamefont
  {Gingras}(2001)}]{Bramwell2001}%
  \BibitemOpen
  \bibfield  {author} {\bibinfo {author} {\bibfnamefont {S.~T.}\ \bibnamefont
  {Bramwell}}\ and\ \bibinfo {author} {\bibfnamefont {M.~J.~P.}\ \bibnamefont
  {Gingras}},\ }\href@noop {} {\bibfield  {journal} {\bibinfo  {journal}
  {Science}\ }\textbf {\bibinfo {volume} {294}},\ \bibinfo {pages} {1495}
  (\bibinfo {year} {2001})}\BibitemShut {NoStop}%
\bibitem [{\citenamefont {Harris}\ \emph {et~al.}(1997)\citenamefont {Harris},
  \citenamefont {Bramwell}, \citenamefont {McMorrow}, \citenamefont {Zeiske},\
  and\ \citenamefont {Godfrey}}]{Harris1997}%
  \BibitemOpen
  \bibfield  {author} {\bibinfo {author} {\bibfnamefont {M.~J.}\ \bibnamefont
  {Harris}}, \bibinfo {author} {\bibfnamefont {S.~T.}\ \bibnamefont
  {Bramwell}}, \bibinfo {author} {\bibfnamefont {D.~F.}\ \bibnamefont
  {McMorrow}}, \bibinfo {author} {\bibfnamefont {T.}~\bibnamefont {Zeiske}}, \
  and\ \bibinfo {author} {\bibfnamefont {K.~W.}\ \bibnamefont {Godfrey}},\
  }\href {\doibase 10.1103/PhysRevLett.79.2554} {\bibfield  {journal} {\bibinfo
   {journal} {Phys. Rev. Lett.}\ }\textbf {\bibinfo {volume} {79}},\ \bibinfo
  {pages} {2554} (\bibinfo {year} {1997})}\BibitemShut {NoStop}%
\bibitem [{\citenamefont {Gardner}\ \emph
  {et~al.}(1999{\natexlab{a}})\citenamefont {Gardner}, \citenamefont {Gaulin},
  \citenamefont {Lee}, \citenamefont {Broholm}, \citenamefont {Raju},\ and\
  \citenamefont {Greedan}}]{GardnerGaulin1999}%
  \BibitemOpen
  \bibfield  {author} {\bibinfo {author} {\bibfnamefont {J.}~\bibnamefont
  {Gardner}}, \bibinfo {author} {\bibfnamefont {B.}~\bibnamefont {Gaulin}},
  \bibinfo {author} {\bibfnamefont {S.-H.}\ \bibnamefont {Lee}}, \bibinfo
  {author} {\bibfnamefont {C.}~\bibnamefont {Broholm}}, \bibinfo {author}
  {\bibfnamefont {N.}~\bibnamefont {Raju}}, \ and\ \bibinfo {author}
  {\bibfnamefont {J.}~\bibnamefont {Greedan}},\ }\href@noop {} {\bibfield
  {journal} {\bibinfo  {journal} {Phys. Rev. Lett.}\ }\textbf {\bibinfo
  {volume} {83}},\ \bibinfo {pages} {211} (\bibinfo {year}
  {1999}{\natexlab{a}})}\BibitemShut {NoStop}%
\bibitem [{\citenamefont {Singh}\ and\ \citenamefont {Lee}(2012)}]{Singh2012}%
  \BibitemOpen
  \bibfield  {author} {\bibinfo {author} {\bibfnamefont {D.~K.}\ \bibnamefont
  {Singh}}\ and\ \bibinfo {author} {\bibfnamefont {Y.~S.}\ \bibnamefont
  {Lee}},\ }\href@noop {} {\bibfield  {journal} {\bibinfo  {journal} {Phys.
  Rev. Lett.}\ }\textbf {\bibinfo {volume} {109}},\ \bibinfo {pages} {247201}
  (\bibinfo {year} {2012})}\BibitemShut {NoStop}%
\bibitem [{\citenamefont {Singh}\ \emph {et~al.}(2008)\citenamefont {Singh},
  \citenamefont {Helton}, \citenamefont {Chu}, \citenamefont {Han},
  \citenamefont {Bonnoit}, \citenamefont {Chang}, \citenamefont {Kang},
  \citenamefont {Lynn},\ and\ \citenamefont {Lee}}]{Singh2008}%
  \BibitemOpen
  \bibfield  {author} {\bibinfo {author} {\bibfnamefont {D.~K.}\ \bibnamefont
  {Singh}}, \bibinfo {author} {\bibfnamefont {J.~S.}\ \bibnamefont {Helton}},
  \bibinfo {author} {\bibfnamefont {S.}~\bibnamefont {Chu}}, \bibinfo {author}
  {\bibfnamefont {T.~H.}\ \bibnamefont {Han}}, \bibinfo {author} {\bibfnamefont
  {C.~J.}\ \bibnamefont {Bonnoit}}, \bibinfo {author} {\bibfnamefont
  {S.}~\bibnamefont {Chang}}, \bibinfo {author} {\bibfnamefont {H.~J.}\
  \bibnamefont {Kang}}, \bibinfo {author} {\bibfnamefont {J.~W.}\ \bibnamefont
  {Lynn}}, \ and\ \bibinfo {author} {\bibfnamefont {Y.~S.}\ \bibnamefont
  {Lee}},\ }\href@noop {} {\bibfield  {journal} {\bibinfo  {journal} {Phys.
  Rev. B}\ }\textbf {\bibinfo {volume} {78}},\ \bibinfo {pages} {220405}
  (\bibinfo {year} {2008})}\BibitemShut {NoStop}%
\bibitem [{\citenamefont {Greedan}\ \emph {et~al.}(1996)\citenamefont
  {Greedan}, \citenamefont {Raju}, \citenamefont {Maignan}, \citenamefont
  {Simon}, \citenamefont {Pedersen}, \citenamefont {Niraimathi}, \citenamefont
  {Gmelin},\ and\ \citenamefont {Subramanian}}]{Greedan1996}%
  \BibitemOpen
  \bibfield  {author} {\bibinfo {author} {\bibfnamefont {J.~E.}\ \bibnamefont
  {Greedan}}, \bibinfo {author} {\bibfnamefont {N.~P.}\ \bibnamefont {Raju}},
  \bibinfo {author} {\bibfnamefont {A.}~\bibnamefont {Maignan}}, \bibinfo
  {author} {\bibfnamefont {C.}~\bibnamefont {Simon}}, \bibinfo {author}
  {\bibfnamefont {J.~S.}\ \bibnamefont {Pedersen}}, \bibinfo {author}
  {\bibfnamefont {A.~M.}\ \bibnamefont {Niraimathi}}, \bibinfo {author}
  {\bibfnamefont {E.}~\bibnamefont {Gmelin}}, \ and\ \bibinfo {author}
  {\bibfnamefont {M.~A.}\ \bibnamefont {Subramanian}},\ }\href {\doibase
  10.1103/PhysRevB.54.7189} {\bibfield  {journal} {\bibinfo  {journal} {Phys.
  Rev. B}\ }\textbf {\bibinfo {volume} {54}},\ \bibinfo {pages} {7189}
  (\bibinfo {year} {1996})}\BibitemShut {NoStop}%
\bibitem [{\citenamefont {Gaulin}\ \emph {et~al.}(1992)\citenamefont {Gaulin},
  \citenamefont {Reimers}, \citenamefont {Mason}, \citenamefont {Greedan},\
  and\ \citenamefont {Tun}}]{Gaulin1992}%
  \BibitemOpen
  \bibfield  {author} {\bibinfo {author} {\bibfnamefont {B.~D.}\ \bibnamefont
  {Gaulin}}, \bibinfo {author} {\bibfnamefont {J.~N.}\ \bibnamefont {Reimers}},
  \bibinfo {author} {\bibfnamefont {T.~E.}\ \bibnamefont {Mason}}, \bibinfo
  {author} {\bibfnamefont {J.~E.}\ \bibnamefont {Greedan}}, \ and\ \bibinfo
  {author} {\bibfnamefont {Z.}~\bibnamefont {Tun}},\ }\href {\doibase
  10.1103/PhysRevLett.69.3244} {\bibfield  {journal} {\bibinfo  {journal}
  {Phys. Rev. Lett.}\ }\textbf {\bibinfo {volume} {69}},\ \bibinfo {pages}
  {3244} (\bibinfo {year} {1992})}\BibitemShut {NoStop}%
\bibitem [{\citenamefont {Greedan}\ \emph {et~al.}(1991)\citenamefont
  {Greedan}, \citenamefont {Reimers}, \citenamefont {Stager},\ and\
  \citenamefont {Penny}}]{Greedan1991}%
  \BibitemOpen
  \bibfield  {author} {\bibinfo {author} {\bibfnamefont {J.~E.}\ \bibnamefont
  {Greedan}}, \bibinfo {author} {\bibfnamefont {J.~N.}\ \bibnamefont
  {Reimers}}, \bibinfo {author} {\bibfnamefont {C.~V.}\ \bibnamefont {Stager}},
  \ and\ \bibinfo {author} {\bibfnamefont {S.~L.}\ \bibnamefont {Penny}},\
  }\href {\doibase 10.1103/PhysRevB.43.5682} {\bibfield  {journal} {\bibinfo
  {journal} {Phys. Rev. B}\ }\textbf {\bibinfo {volume} {43}},\ \bibinfo
  {pages} {5682} (\bibinfo {year} {1991})}\BibitemShut {NoStop}%
\bibitem [{\citenamefont {Gingras}\ \emph {et~al.}(1997)\citenamefont
  {Gingras}, \citenamefont {Stager}, \citenamefont {Raju}, \citenamefont
  {Gaulin},\ and\ \citenamefont {Greedan}}]{Gingras1997}%
  \BibitemOpen
  \bibfield  {author} {\bibinfo {author} {\bibfnamefont {M.~J.~P.}\
  \bibnamefont {Gingras}}, \bibinfo {author} {\bibfnamefont {C.~V.}\
  \bibnamefont {Stager}}, \bibinfo {author} {\bibfnamefont {N.~P.}\
  \bibnamefont {Raju}}, \bibinfo {author} {\bibfnamefont {B.~D.}\ \bibnamefont
  {Gaulin}}, \ and\ \bibinfo {author} {\bibfnamefont {J.~E.}\ \bibnamefont
  {Greedan}},\ }\href {\doibase 10.1103/PhysRevLett.78.947} {\bibfield
  {journal} {\bibinfo  {journal} {Phys. Rev. Lett.}\ }\textbf {\bibinfo
  {volume} {78}},\ \bibinfo {pages} {947} (\bibinfo {year} {1997})}\BibitemShut
  {NoStop}%
\bibitem [{\citenamefont {Kanada}\ \emph {et~al.}(1999)\citenamefont {Kanada},
  \citenamefont {Yasui}, \citenamefont {Ito}, \citenamefont {Harashina},
  \citenamefont {Sato}, \citenamefont {Okumura},\ and\ \citenamefont
  {Kakurai}}]{Kanada1999}%
  \BibitemOpen
  \bibfield  {author} {\bibinfo {author} {\bibfnamefont {M.}~\bibnamefont
  {Kanada}}, \bibinfo {author} {\bibfnamefont {Y.}~\bibnamefont {Yasui}},
  \bibinfo {author} {\bibfnamefont {M.}~\bibnamefont {Ito}}, \bibinfo {author}
  {\bibfnamefont {H.}~\bibnamefont {Harashina}}, \bibinfo {author}
  {\bibfnamefont {M.}~\bibnamefont {Sato}}, \bibinfo {author} {\bibfnamefont
  {H.}~\bibnamefont {Okumura}}, \ and\ \bibinfo {author} {\bibfnamefont
  {K.}~\bibnamefont {Kakurai}},\ }\href {\doibase 10.1143/JPSJ.68.3802}
  {\bibfield  {journal} {\bibinfo  {journal} {J. Phys. Soc. Jpn.}\ }\textbf
  {\bibinfo {volume} {68}},\ \bibinfo {pages} {3802} (\bibinfo {year}
  {1999})}\BibitemShut {NoStop}%
\bibitem [{\citenamefont {Mirebeau}\ \emph {et~al.}(2007)\citenamefont
  {Mirebeau}, \citenamefont {Bonville},\ and\ \citenamefont
  {Hennion}}]{Mirebeau2007}%
  \BibitemOpen
  \bibfield  {author} {\bibinfo {author} {\bibfnamefont {I.}~\bibnamefont
  {Mirebeau}}, \bibinfo {author} {\bibfnamefont {P.}~\bibnamefont {Bonville}},
  \ and\ \bibinfo {author} {\bibfnamefont {M.}~\bibnamefont {Hennion}},\
  }\href@noop {} {\bibfield  {journal} {\bibinfo  {journal} {Phys. Rev. B}\
  }\textbf {\bibinfo {volume} {76}},\ \bibinfo {pages} {184436} (\bibinfo
  {year} {2007})}\BibitemShut {NoStop}%
\bibitem [{\citenamefont {Gardner}\ \emph
  {et~al.}(1999{\natexlab{b}})\citenamefont {Gardner}, \citenamefont
  {Dunsiger}, \citenamefont {Gaulin}, \citenamefont {Gingras}, \citenamefont
  {Greedan}, \citenamefont {Kiefl}, \citenamefont {Lumsden}, \citenamefont
  {MacFarlane}, \citenamefont {Raju}, \citenamefont {Sonier}, \citenamefont
  {Swainson},\ and\ \citenamefont {Tun}}]{Gardner1999}%
  \BibitemOpen
  \bibfield  {author} {\bibinfo {author} {\bibfnamefont {J.~S.}\ \bibnamefont
  {Gardner}}, \bibinfo {author} {\bibfnamefont {S.~R.}\ \bibnamefont
  {Dunsiger}}, \bibinfo {author} {\bibfnamefont {B.~D.}\ \bibnamefont
  {Gaulin}}, \bibinfo {author} {\bibfnamefont {M.~J.~P.}\ \bibnamefont
  {Gingras}}, \bibinfo {author} {\bibfnamefont {J.~E.}\ \bibnamefont
  {Greedan}}, \bibinfo {author} {\bibfnamefont {R.~F.}\ \bibnamefont {Kiefl}},
  \bibinfo {author} {\bibfnamefont {M.~D.}\ \bibnamefont {Lumsden}}, \bibinfo
  {author} {\bibfnamefont {W.~A.}\ \bibnamefont {MacFarlane}}, \bibinfo
  {author} {\bibfnamefont {N.~P.}\ \bibnamefont {Raju}}, \bibinfo {author}
  {\bibfnamefont {J.~E.}\ \bibnamefont {Sonier}}, \bibinfo {author}
  {\bibfnamefont {I.}~\bibnamefont {Swainson}}, \ and\ \bibinfo {author}
  {\bibfnamefont {Z.}~\bibnamefont {Tun}},\ }\href@noop {} {\bibfield
  {journal} {\bibinfo  {journal} {Phys. Rev. Lett.}\ }\textbf {\bibinfo
  {volume} {82}},\ \bibinfo {pages} {1012} (\bibinfo {year}
  {1999}{\natexlab{b}})}\BibitemShut {NoStop}%
\bibitem [{\citenamefont {Gardner}\ \emph {et~al.}(2001)\citenamefont
  {Gardner}, \citenamefont {Gaulin}, \citenamefont {Berlinsky}, \citenamefont
  {Waldron}, \citenamefont {Dunsiger}, \citenamefont {Raju},\ and\
  \citenamefont {Greedan}}]{Gardner2001}%
  \BibitemOpen
  \bibfield  {author} {\bibinfo {author} {\bibfnamefont {J.~S.}\ \bibnamefont
  {Gardner}}, \bibinfo {author} {\bibfnamefont {B.~D.}\ \bibnamefont {Gaulin}},
  \bibinfo {author} {\bibfnamefont {A.~J.}\ \bibnamefont {Berlinsky}}, \bibinfo
  {author} {\bibfnamefont {P.}~\bibnamefont {Waldron}}, \bibinfo {author}
  {\bibfnamefont {S.~R.}\ \bibnamefont {Dunsiger}}, \bibinfo {author}
  {\bibfnamefont {N.~P.}\ \bibnamefont {Raju}}, \ and\ \bibinfo {author}
  {\bibfnamefont {J.~E.}\ \bibnamefont {Greedan}},\ }\href@noop {} {\bibfield
  {journal} {\bibinfo  {journal} {Phys. Rev. B}\ }\textbf {\bibinfo {volume}
  {64}},\ \bibinfo {pages} {224416} (\bibinfo {year} {2001})}\BibitemShut
  {NoStop}%
\bibitem [{\citenamefont {Gardner}\ \emph {et~al.}(2003)\citenamefont
  {Gardner}, \citenamefont {Keren}, \citenamefont {Ehlers}, \citenamefont
  {Stock}, \citenamefont {Segal}, \citenamefont {Roper}, \citenamefont
  {F\aa{}k}, \citenamefont {Stone}, \citenamefont {Hammar}, \citenamefont
  {Reich},\ and\ \citenamefont {Gaulin}}]{Gardner2003}%
  \BibitemOpen
  \bibfield  {author} {\bibinfo {author} {\bibfnamefont {J.~S.}\ \bibnamefont
  {Gardner}}, \bibinfo {author} {\bibfnamefont {A.}~\bibnamefont {Keren}},
  \bibinfo {author} {\bibfnamefont {G.}~\bibnamefont {Ehlers}}, \bibinfo
  {author} {\bibfnamefont {C.}~\bibnamefont {Stock}}, \bibinfo {author}
  {\bibfnamefont {E.}~\bibnamefont {Segal}}, \bibinfo {author} {\bibfnamefont
  {J.~M.}\ \bibnamefont {Roper}}, \bibinfo {author} {\bibfnamefont
  {B.}~\bibnamefont {F\aa{}k}}, \bibinfo {author} {\bibfnamefont {M.~B.}\
  \bibnamefont {Stone}}, \bibinfo {author} {\bibfnamefont {P.~R.}\ \bibnamefont
  {Hammar}}, \bibinfo {author} {\bibfnamefont {D.~H.}\ \bibnamefont {Reich}}, \
  and\ \bibinfo {author} {\bibfnamefont {B.~D.}\ \bibnamefont {Gaulin}},\
  }\href@noop {} {\bibfield  {journal} {\bibinfo  {journal} {Phys. Rev. B}\
  }\textbf {\bibinfo {volume} {68}},\ \bibinfo {pages} {180401} (\bibinfo
  {year} {2003})}\BibitemShut {NoStop}%
\bibitem [{\citenamefont {den Hertog}\ and\ \citenamefont
  {Gingras}(2000)}]{Hertog2000}%
  \BibitemOpen
  \bibfield  {author} {\bibinfo {author} {\bibfnamefont {B.~C.}\ \bibnamefont
  {den Hertog}}\ and\ \bibinfo {author} {\bibfnamefont {M.~J.~P.}\ \bibnamefont
  {Gingras}},\ }\href {\doibase 10.1103/PhysRevLett.84.3430} {\bibfield
  {journal} {\bibinfo  {journal} {Phys. Rev. Lett.}\ }\textbf {\bibinfo
  {volume} {84}},\ \bibinfo {pages} {3430} (\bibinfo {year}
  {2000})}\BibitemShut {NoStop}%
\bibitem [{\citenamefont {Ross}\ \emph {et~al.}(2011)\citenamefont {Ross},
  \citenamefont {Savary}, \citenamefont {Gaulin},\ and\ \citenamefont
  {Balents}}]{Ross2011}%
  \BibitemOpen
  \bibfield  {author} {\bibinfo {author} {\bibfnamefont {K.~A.}\ \bibnamefont
  {Ross}}, \bibinfo {author} {\bibfnamefont {L.}~\bibnamefont {Savary}},
  \bibinfo {author} {\bibfnamefont {B.~D.}\ \bibnamefont {Gaulin}}, \ and\
  \bibinfo {author} {\bibfnamefont {L.}~\bibnamefont {Balents}},\ }\href
  {http://link.aps.org/doi/10.1103/PhysRevX.1.021002} {\bibfield  {journal}
  {\bibinfo  {journal} {Phys. Rev. X}\ }\textbf {\bibinfo {volume} {1}},\
  \bibinfo {pages} {021002} (\bibinfo {year} {2011})}\BibitemShut {NoStop}%
\bibitem [{\citenamefont {Kimura}\ \emph {et~al.}(2013)\citenamefont {Kimura},
  \citenamefont {Nakatsuji}, \citenamefont {Wen}, \citenamefont {Broholm},
  \citenamefont {Stone}, \citenamefont {Nishibori},\ and\ \citenamefont
  {Sawa}}]{Kimura2013}%
  \BibitemOpen
  \bibfield  {author} {\bibinfo {author} {\bibfnamefont {K.}~\bibnamefont
  {Kimura}}, \bibinfo {author} {\bibfnamefont {S.}~\bibnamefont {Nakatsuji}},
  \bibinfo {author} {\bibfnamefont {J.-J.}\ \bibnamefont {Wen}}, \bibinfo
  {author} {\bibfnamefont {C.}~\bibnamefont {Broholm}}, \bibinfo {author}
  {\bibfnamefont {M.~B.}\ \bibnamefont {Stone}}, \bibinfo {author}
  {\bibfnamefont {E.}~\bibnamefont {Nishibori}}, \ and\ \bibinfo {author}
  {\bibfnamefont {H.}~\bibnamefont {Sawa}},\ }\href@noop {} {\bibfield
  {journal} {\bibinfo  {journal} {Nature Commun.}\ }\textbf {\bibinfo {volume}
  {4}} (\bibinfo {year} {2013})}\BibitemShut {NoStop}%
\bibitem [{\citenamefont {Champion}\ \emph {et~al.}(2003)\citenamefont
  {Champion}, \citenamefont {Harris}, \citenamefont {Holdsworth}, \citenamefont
  {Wills}, \citenamefont {Balakrishnan}, \citenamefont {Bramwell},
  \citenamefont {{\v{C}}i{\v{z}}m\'ar}, \citenamefont {Fennell}, \citenamefont
  {Gardner}, \citenamefont {Lago}, \citenamefont {McMorrow}, \citenamefont
  {Orend\'a\ifmmode~\check{c}\else \v{c}\fi{}}, \citenamefont
  {Orend\'a\ifmmode~\check{c}\else \v{c}\fi{}ov\'a}, \citenamefont {Paul},
  \citenamefont {Smith}, \citenamefont {Telling},\ and\ \citenamefont
  {Wildes}}]{Champion2003}%
  \BibitemOpen
  \bibfield  {author} {\bibinfo {author} {\bibfnamefont {J.~D.~M.}\
  \bibnamefont {Champion}}, \bibinfo {author} {\bibfnamefont {M.~J.}\
  \bibnamefont {Harris}}, \bibinfo {author} {\bibfnamefont {P.~C.~W.}\
  \bibnamefont {Holdsworth}}, \bibinfo {author} {\bibfnamefont {A.~S.}\
  \bibnamefont {Wills}}, \bibinfo {author} {\bibfnamefont {G.}~\bibnamefont
  {Balakrishnan}}, \bibinfo {author} {\bibfnamefont {S.~T.}\ \bibnamefont
  {Bramwell}}, \bibinfo {author} {\bibfnamefont {E.}~\bibnamefont
  {{\v{C}}i{\v{z}}m\'ar}}, \bibinfo {author} {\bibfnamefont {T.}~\bibnamefont
  {Fennell}}, \bibinfo {author} {\bibfnamefont {J.~S.}\ \bibnamefont
  {Gardner}}, \bibinfo {author} {\bibfnamefont {J.}~\bibnamefont {Lago}},
  \bibinfo {author} {\bibfnamefont {D.~F.}\ \bibnamefont {McMorrow}}, \bibinfo
  {author} {\bibfnamefont {M.}~\bibnamefont {Orend\'a\ifmmode~\check{c}\else
  \v{c}\fi{}}}, \bibinfo {author} {\bibfnamefont {A.}~\bibnamefont
  {Orend\'a\ifmmode~\check{c}\else \v{c}\fi{}ov\'a}}, \bibinfo {author}
  {\bibfnamefont {D.~M.}\ \bibnamefont {Paul}}, \bibinfo {author}
  {\bibfnamefont {R.~I.}\ \bibnamefont {Smith}}, \bibinfo {author}
  {\bibfnamefont {M.~T.~F.}\ \bibnamefont {Telling}}, \ and\ \bibinfo {author}
  {\bibfnamefont {A.}~\bibnamefont {Wildes}},\ }\href {\doibase
  10.1103/PhysRevB.68.020401} {\bibfield  {journal} {\bibinfo  {journal} {Phys.
  Rev. B}\ }\textbf {\bibinfo {volume} {68}},\ \bibinfo {pages} {020401}
  (\bibinfo {year} {2003})}\BibitemShut {NoStop}%
\bibitem [{\citenamefont {Savary}\ \emph {et~al.}(2012)\citenamefont {Savary},
  \citenamefont {Ross}, \citenamefont {Gaulin}, \citenamefont {Ruff},\ and\
  \citenamefont {Balents}}]{Savary2012}%
  \BibitemOpen
  \bibfield  {author} {\bibinfo {author} {\bibfnamefont {L.}~\bibnamefont
  {Savary}}, \bibinfo {author} {\bibfnamefont {K.~A.}\ \bibnamefont {Ross}},
  \bibinfo {author} {\bibfnamefont {B.~D.}\ \bibnamefont {Gaulin}}, \bibinfo
  {author} {\bibfnamefont {J.~P.~C.}\ \bibnamefont {Ruff}}, \ and\ \bibinfo
  {author} {\bibfnamefont {L.}~\bibnamefont {Balents}},\ }\href
  {http://link.aps.org/doi/10.1103/PhysRevLett.109.167201} {\bibfield
  {journal} {\bibinfo  {journal} {Phys. Rev. Lett.}\ }\textbf {\bibinfo
  {volume} {109}},\ \bibinfo {pages} {167201} (\bibinfo {year}
  {2012})}\BibitemShut {NoStop}%
\bibitem [{\citenamefont {Zhitomirsky}\ \emph {et~al.}(2012)\citenamefont
  {Zhitomirsky}, \citenamefont {Gvozdikova}, \citenamefont {Holdsworth},\ and\
  \citenamefont {Moessner}}]{Zhitomirsky2012}%
  \BibitemOpen
  \bibfield  {author} {\bibinfo {author} {\bibfnamefont {M.~E.}\ \bibnamefont
  {Zhitomirsky}}, \bibinfo {author} {\bibfnamefont {M.~V.}\ \bibnamefont
  {Gvozdikova}}, \bibinfo {author} {\bibfnamefont {P.~C.~W.}\ \bibnamefont
  {Holdsworth}}, \ and\ \bibinfo {author} {\bibfnamefont {R.}~\bibnamefont
  {Moessner}},\ }\href {http://link.aps.org/doi/10.1103/PhysRevLett.109.077204}
  {\bibfield  {journal} {\bibinfo  {journal} {Phys. Rev. Lett.}\ }\textbf
  {\bibinfo {volume} {109}},\ \bibinfo {pages} {077204} (\bibinfo {year}
  {2012})}\BibitemShut {NoStop}%
\bibitem [{\citenamefont {Gardner}\ \emph {et~al.}(2010)\citenamefont
  {Gardner}, \citenamefont {Gingras},\ and\ \citenamefont
  {Greedan}}]{Gardner2010}%
  \BibitemOpen
  \bibfield  {author} {\bibinfo {author} {\bibfnamefont {J.~S.}\ \bibnamefont
  {Gardner}}, \bibinfo {author} {\bibfnamefont {M.~J.~P.}\ \bibnamefont
  {Gingras}}, \ and\ \bibinfo {author} {\bibfnamefont {J.~E.}\ \bibnamefont
  {Greedan}},\ }\href {http://link.aps.org/doi/10.1103/RevModPhys.82.53}
  {\bibfield  {journal} {\bibinfo  {journal} {Rev. Mod. Phys.}\ }\textbf
  {\bibinfo {volume} {82}},\ \bibinfo {pages} {53} (\bibinfo {year}
  {2010})}\BibitemShut {NoStop}%
\bibitem [{\citenamefont {Zinkin}\ \emph {et~al.}(1996)\citenamefont {Zinkin},
  \citenamefont {Harris}, \citenamefont {Tun}, \citenamefont {Cowley},\ and\
  \citenamefont {Wanklyn}}]{Zinkin1996}%
  \BibitemOpen
  \bibfield  {author} {\bibinfo {author} {\bibfnamefont {M.~P.}\ \bibnamefont
  {Zinkin}}, \bibinfo {author} {\bibfnamefont {M.~J.}\ \bibnamefont {Harris}},
  \bibinfo {author} {\bibfnamefont {Z.}~\bibnamefont {Tun}}, \bibinfo {author}
  {\bibfnamefont {R.~A.}\ \bibnamefont {Cowley}}, \ and\ \bibinfo {author}
  {\bibfnamefont {B.~M.}\ \bibnamefont {Wanklyn}},\ }\href {\doibase
  10.1088/0953-8984/8/2/007} {\bibfield  {journal} {\bibinfo  {journal} {J.
  Phys.: Condens. Matter}\ }\textbf {\bibinfo {volume} {8}},\ \bibinfo {pages}
  {193} (\bibinfo {year} {1996})}\BibitemShut {NoStop}%
\bibitem [{\citenamefont {Rosenkranz}\ \emph {et~al.}(2000)\citenamefont
  {Rosenkranz}, \citenamefont {Ramirez}, \citenamefont {Hayashi}, \citenamefont
  {Cava}, \citenamefont {Siddharthan},\ and\ \citenamefont
  {Shastry}}]{Rosenkranz2000}%
  \BibitemOpen
  \bibfield  {author} {\bibinfo {author} {\bibfnamefont {S.}~\bibnamefont
  {Rosenkranz}}, \bibinfo {author} {\bibfnamefont {A.~P.}\ \bibnamefont
  {Ramirez}}, \bibinfo {author} {\bibfnamefont {A.}~\bibnamefont {Hayashi}},
  \bibinfo {author} {\bibfnamefont {R.~J.}\ \bibnamefont {Cava}}, \bibinfo
  {author} {\bibfnamefont {R.}~\bibnamefont {Siddharthan}}, \ and\ \bibinfo
  {author} {\bibfnamefont {B.~S.}\ \bibnamefont {Shastry}},\ }\href {\doibase
  10.1063/1.372565} {\bibfield  {journal} {\bibinfo  {journal} {J. Appl.
  Phys.}\ }\textbf {\bibinfo {volume} {87}},\ \bibinfo {pages} {5914} (\bibinfo
  {year} {2000})}\BibitemShut {NoStop}%
\bibitem [{\citenamefont {Gardner}\ and\ \citenamefont
  {Ehlers}(2009)}]{Gardner2009}%
  \BibitemOpen
  \bibfield  {author} {\bibinfo {author} {\bibfnamefont {J.~S.}\ \bibnamefont
  {Gardner}}\ and\ \bibinfo {author} {\bibfnamefont {G.}~\bibnamefont
  {Ehlers}},\ }\href {http://dx.doi.org/10.1088/0953-8984/21/43/436004}
  {\bibfield  {journal} {\bibinfo  {journal} {J. Phys.: Condens. Matter}\
  }\textbf {\bibinfo {volume} {21}},\ \bibinfo {pages} {436004} (\bibinfo
  {year} {2009})}\BibitemShut {NoStop}%
\bibitem [{\citenamefont {Bertin}\ \emph {et~al.}(2012)\citenamefont {Bertin},
  \citenamefont {Chapuis}, \citenamefont {Dalmas~de Reotier},\ and\
  \citenamefont {Yaouanc}}]{Bertin2012}%
  \BibitemOpen
  \bibfield  {author} {\bibinfo {author} {\bibfnamefont {A.}~\bibnamefont
  {Bertin}}, \bibinfo {author} {\bibfnamefont {Y.}~\bibnamefont {Chapuis}},
  \bibinfo {author} {\bibfnamefont {P.}~\bibnamefont {Dalmas~de Reotier}}, \
  and\ \bibinfo {author} {\bibfnamefont {A.}~\bibnamefont {Yaouanc}},\ }\href
  {http://dx.doi.org/10.1088/0953-8984/24/25/256003} {\bibfield  {journal}
  {\bibinfo  {journal} {J. Phys.: Condens. Matter}\ }\textbf {\bibinfo {volume}
  {24}},\ \bibinfo {pages} {256003} (\bibinfo {year} {2012})}\BibitemShut
  {NoStop}%
\bibitem [{\citenamefont {Gingras}\ \emph {et~al.}(2000)\citenamefont
  {Gingras}, \citenamefont {den Hertog}, \citenamefont {Faucher}, \citenamefont
  {Gardner}, \citenamefont {Dunsiger}, \citenamefont {Chang}, \citenamefont
  {Gaulin}, \citenamefont {Raju},\ and\ \citenamefont {Greedan}}]{Gingras2000}%
  \BibitemOpen
  \bibfield  {author} {\bibinfo {author} {\bibfnamefont {M.~J.~P.}\
  \bibnamefont {Gingras}}, \bibinfo {author} {\bibfnamefont {B.~C.}\
  \bibnamefont {den Hertog}}, \bibinfo {author} {\bibfnamefont
  {M.}~\bibnamefont {Faucher}}, \bibinfo {author} {\bibfnamefont {J.~S.}\
  \bibnamefont {Gardner}}, \bibinfo {author} {\bibfnamefont {S.~R.}\
  \bibnamefont {Dunsiger}}, \bibinfo {author} {\bibfnamefont {L.~J.}\
  \bibnamefont {Chang}}, \bibinfo {author} {\bibfnamefont {B.~D.}\ \bibnamefont
  {Gaulin}}, \bibinfo {author} {\bibfnamefont {N.~P.}\ \bibnamefont {Raju}}, \
  and\ \bibinfo {author} {\bibfnamefont {J.~E.}\ \bibnamefont {Greedan}},\
  }\href@noop {} {\bibfield  {journal} {\bibinfo  {journal} {Phys. Rev. B}\
  }\textbf {\bibinfo {volume} {62}},\ \bibinfo {pages} {6496} (\bibinfo {year}
  {2000})}\BibitemShut {NoStop}%
\bibitem [{\citenamefont {Princep}\ \emph {et~al.}(2013)\citenamefont
  {Princep}, \citenamefont {Prabhakaran}, \citenamefont {Boothroyd},\ and\
  \citenamefont {Adroja}}]{Princep2013}%
  \BibitemOpen
  \bibfield  {author} {\bibinfo {author} {\bibfnamefont {A.~J.}\ \bibnamefont
  {Princep}}, \bibinfo {author} {\bibfnamefont {D.}~\bibnamefont
  {Prabhakaran}}, \bibinfo {author} {\bibfnamefont {A.~T.}\ \bibnamefont
  {Boothroyd}}, \ and\ \bibinfo {author} {\bibfnamefont {D.~T.}\ \bibnamefont
  {Adroja}},\ }\href@noop {} {\bibfield  {journal} {\bibinfo  {journal} {Phys.
  Rev. B}\ }\textbf {\bibinfo {volume} {88}},\ \bibinfo {pages} {104421}
  (\bibinfo {year} {2013})}\BibitemShut {NoStop}%
\bibitem [{\citenamefont {Matsuhira}\ \emph {et~al.}(2002)\citenamefont
  {Matsuhira}, \citenamefont {Hinatsu}, \citenamefont {Tenya}, \citenamefont
  {Amitsuka},\ and\ \citenamefont {Sakakibara}}]{Matsuhira2002}%
  \BibitemOpen
  \bibfield  {author} {\bibinfo {author} {\bibfnamefont {K.}~\bibnamefont
  {Matsuhira}}, \bibinfo {author} {\bibfnamefont {Y.}~\bibnamefont {Hinatsu}},
  \bibinfo {author} {\bibfnamefont {K.}~\bibnamefont {Tenya}}, \bibinfo
  {author} {\bibfnamefont {H.}~\bibnamefont {Amitsuka}}, \ and\ \bibinfo
  {author} {\bibfnamefont {T.}~\bibnamefont {Sakakibara}},\ }\href {\doibase
  10.1143/JPSJ.71.1576} {\bibfield  {journal} {\bibinfo  {journal} {J. Phys.
  Soc. Jpn.}\ }\textbf {\bibinfo {volume} {71}},\ \bibinfo {pages} {1576}
  (\bibinfo {year} {2002})}\BibitemShut {NoStop}%
\bibitem [{\citenamefont {Mirebeau}\ \emph {et~al.}(2005)\citenamefont
  {Mirebeau}, \citenamefont {Apetrei}, \citenamefont {Rodriguez-Carvajal},
  \citenamefont {Bonville}, \citenamefont {Forget}, \citenamefont {Colson},
  \citenamefont {Glazkov}, \citenamefont {Sanchez}, \citenamefont {Isnard},\
  and\ \citenamefont {Suard}}]{Mirebeau2005}%
  \BibitemOpen
  \bibfield  {author} {\bibinfo {author} {\bibfnamefont {I.}~\bibnamefont
  {Mirebeau}}, \bibinfo {author} {\bibfnamefont {A.}~\bibnamefont {Apetrei}},
  \bibinfo {author} {\bibfnamefont {J.}~\bibnamefont {Rodriguez-Carvajal}},
  \bibinfo {author} {\bibfnamefont {P.}~\bibnamefont {Bonville}}, \bibinfo
  {author} {\bibfnamefont {A.}~\bibnamefont {Forget}}, \bibinfo {author}
  {\bibfnamefont {D.}~\bibnamefont {Colson}}, \bibinfo {author} {\bibfnamefont
  {V.}~\bibnamefont {Glazkov}}, \bibinfo {author} {\bibfnamefont {J.~P.}\
  \bibnamefont {Sanchez}}, \bibinfo {author} {\bibfnamefont {O.}~\bibnamefont
  {Isnard}}, \ and\ \bibinfo {author} {\bibfnamefont {E.}~\bibnamefont
  {Suard}},\ }\href@noop {} {\bibfield  {journal} {\bibinfo  {journal} {Phys.
  Rev. Lett.}\ }\textbf {\bibinfo {volume} {94}},\ \bibinfo {pages} {246402}
  (\bibinfo {year} {2005})}\BibitemShut {NoStop}%
\bibitem [{\citenamefont {Hamaguchi}\ \emph {et~al.}(2004)\citenamefont
  {Hamaguchi}, \citenamefont {Matsushita}, \citenamefont {Wada}, \citenamefont
  {Yasui},\ and\ \citenamefont {Sato}}]{Hamaguchi2004}%
  \BibitemOpen
  \bibfield  {author} {\bibinfo {author} {\bibfnamefont {N.}~\bibnamefont
  {Hamaguchi}}, \bibinfo {author} {\bibfnamefont {T.}~\bibnamefont
  {Matsushita}}, \bibinfo {author} {\bibfnamefont {N.}~\bibnamefont {Wada}},
  \bibinfo {author} {\bibfnamefont {Y.}~\bibnamefont {Yasui}}, \ and\ \bibinfo
  {author} {\bibfnamefont {M.}~\bibnamefont {Sato}},\ }\href
  {http://link.aps.org/doi/10.1103/PhysRevB.69.132413} {\bibfield  {journal}
  {\bibinfo  {journal} {Phys. Rev. B}\ }\textbf {\bibinfo {volume} {69}},\
  \bibinfo {pages} {132413} (\bibinfo {year} {2004})}\BibitemShut {NoStop}%
\bibitem [{\citenamefont {Yasui}\ \emph {et~al.}(2002)\citenamefont {Yasui},
  \citenamefont {Kanada}, \citenamefont {Ito}, \citenamefont {Harashina},
  \citenamefont {Sato}, \citenamefont {Okumura}, \citenamefont {Kakurai},\ and\
  \citenamefont {Kadowaki}}]{Yasui2002}%
  \BibitemOpen
  \bibfield  {author} {\bibinfo {author} {\bibfnamefont {Y.}~\bibnamefont
  {Yasui}}, \bibinfo {author} {\bibfnamefont {M.}~\bibnamefont {Kanada}},
  \bibinfo {author} {\bibfnamefont {M.}~\bibnamefont {Ito}}, \bibinfo {author}
  {\bibfnamefont {H.}~\bibnamefont {Harashina}}, \bibinfo {author}
  {\bibfnamefont {M.}~\bibnamefont {Sato}}, \bibinfo {author} {\bibfnamefont
  {H.}~\bibnamefont {Okumura}}, \bibinfo {author} {\bibfnamefont
  {K.}~\bibnamefont {Kakurai}}, \ and\ \bibinfo {author} {\bibfnamefont
  {H.}~\bibnamefont {Kadowaki}},\ }\href@noop {} {\bibfield  {journal}
  {\bibinfo  {journal} {J. Phys. Soc. Jpn.}\ }\textbf {\bibinfo {volume}
  {71}},\ \bibinfo {pages} {599} (\bibinfo {year} {2002})}\BibitemShut
  {NoStop}%
\bibitem [{\citenamefont {Lhotel}\ \emph {et~al.}(2012)\citenamefont {Lhotel},
  \citenamefont {Paulsen}, \citenamefont {de~R\'eotier}, \citenamefont
  {Yaouanc}, \citenamefont {Marin},\ and\ \citenamefont
  {Vanishri}}]{Lhotel2012}%
  \BibitemOpen
  \bibfield  {author} {\bibinfo {author} {\bibfnamefont {E.}~\bibnamefont
  {Lhotel}}, \bibinfo {author} {\bibfnamefont {C.}~\bibnamefont {Paulsen}},
  \bibinfo {author} {\bibfnamefont {P.~D.}\ \bibnamefont {de~R\'eotier}},
  \bibinfo {author} {\bibfnamefont {A.}~\bibnamefont {Yaouanc}}, \bibinfo
  {author} {\bibfnamefont {C.}~\bibnamefont {Marin}}, \ and\ \bibinfo {author}
  {\bibfnamefont {S.}~\bibnamefont {Vanishri}},\ }\href {\doibase
  10.1103/PhysRevB.86.020410} {\bibfield  {journal} {\bibinfo  {journal} {Phys.
  Rev. B}\ }\textbf {\bibinfo {volume} {86}},\ \bibinfo {pages} {020410}
  (\bibinfo {year} {2012})}\BibitemShut {NoStop}%
\bibitem [{\citenamefont {Fritsch}\ \emph
  {et~al.}(2013{\natexlab{a}})\citenamefont {Fritsch}, \citenamefont {Ross},
  \citenamefont {Qiu}, \citenamefont {Copley}, \citenamefont {Guidi},
  \citenamefont {Bewley}, \citenamefont {Dabkowska},\ and\ \citenamefont
  {Gaulin}}]{Fritsch2013}%
  \BibitemOpen
  \bibfield  {author} {\bibinfo {author} {\bibfnamefont {K.}~\bibnamefont
  {Fritsch}}, \bibinfo {author} {\bibfnamefont {K.~A.}\ \bibnamefont {Ross}},
  \bibinfo {author} {\bibfnamefont {Y.}~\bibnamefont {Qiu}}, \bibinfo {author}
  {\bibfnamefont {J.~R.~D.}\ \bibnamefont {Copley}}, \bibinfo {author}
  {\bibfnamefont {T.}~\bibnamefont {Guidi}}, \bibinfo {author} {\bibfnamefont
  {R.~I.}\ \bibnamefont {Bewley}}, \bibinfo {author} {\bibfnamefont {H.~A.}\
  \bibnamefont {Dabkowska}}, \ and\ \bibinfo {author} {\bibfnamefont {B.~D.}\
  \bibnamefont {Gaulin}},\ }\href {\doibase 10.1103/PhysRevB.87.094410}
  {\bibfield  {journal} {\bibinfo  {journal} {Phys. Rev. B}\ }\textbf {\bibinfo
  {volume} {87}},\ \bibinfo {pages} {094410} (\bibinfo {year}
  {2013}{\natexlab{a}})}\BibitemShut {NoStop}%
\bibitem [{\citenamefont {Fennell}\ \emph {et~al.}(2012)\citenamefont
  {Fennell}, \citenamefont {Kenzelmann}, \citenamefont {Roessli}, \citenamefont
  {Haas},\ and\ \citenamefont {Cava}}]{Fennell2012}%
  \BibitemOpen
  \bibfield  {author} {\bibinfo {author} {\bibfnamefont {T.}~\bibnamefont
  {Fennell}}, \bibinfo {author} {\bibfnamefont {M.}~\bibnamefont {Kenzelmann}},
  \bibinfo {author} {\bibfnamefont {B.}~\bibnamefont {Roessli}}, \bibinfo
  {author} {\bibfnamefont {M.~K.}\ \bibnamefont {Haas}}, \ and\ \bibinfo
  {author} {\bibfnamefont {R.~J.}\ \bibnamefont {Cava}},\ }\href@noop {}
  {\bibfield  {journal} {\bibinfo  {journal} {Phys. Rev. Lett.}\ }\textbf
  {\bibinfo {volume} {109}},\ \bibinfo {pages} {017201} (\bibinfo {year}
  {2012})}\BibitemShut {NoStop}%
\bibitem [{\citenamefont {Petit}\ \emph {et~al.}(2012)\citenamefont {Petit},
  \citenamefont {Bonville}, \citenamefont {Robert}, \citenamefont {Decorse},\
  and\ \citenamefont {Mirebeau}}]{Petit2012}%
  \BibitemOpen
  \bibfield  {author} {\bibinfo {author} {\bibfnamefont {S.}~\bibnamefont
  {Petit}}, \bibinfo {author} {\bibfnamefont {P.}~\bibnamefont {Bonville}},
  \bibinfo {author} {\bibfnamefont {J.}~\bibnamefont {Robert}}, \bibinfo
  {author} {\bibfnamefont {C.}~\bibnamefont {Decorse}}, \ and\ \bibinfo
  {author} {\bibfnamefont {I.}~\bibnamefont {Mirebeau}},\ }\href
  {http://link.aps.org/doi/10.1103/PhysRevB.86.174403} {\bibfield  {journal}
  {\bibinfo  {journal} {Phys. Rev. B}\ }\textbf {\bibinfo {volume} {86}},\
  \bibinfo {pages} {174403} (\bibinfo {year} {2012})}\BibitemShut {NoStop}%
\bibitem [{\citenamefont {Fritsch}\ \emph
  {et~al.}(2013{\natexlab{b}})\citenamefont {Fritsch}, \citenamefont
  {Kermarrec}, \citenamefont {Ross}, \citenamefont {Qiu}, \citenamefont
  {Copley}, \citenamefont {Pomaranski}, \citenamefont {Kycia}, \citenamefont
  {Dabkowska},\ and\ \citenamefont {Gaulin}}]{Fritscharxiv}%
  \BibitemOpen
  \bibfield  {author} {\bibinfo {author} {\bibfnamefont {K.}~\bibnamefont
  {Fritsch}}, \bibinfo {author} {\bibfnamefont {E.}~\bibnamefont {Kermarrec}},
  \bibinfo {author} {\bibfnamefont {K.~A.}\ \bibnamefont {Ross}}, \bibinfo
  {author} {\bibfnamefont {Y.}~\bibnamefont {Qiu}}, \bibinfo {author}
  {\bibfnamefont {J.~R.~D.}\ \bibnamefont {Copley}}, \bibinfo {author}
  {\bibfnamefont {D.}~\bibnamefont {Pomaranski}}, \bibinfo {author}
  {\bibfnamefont {J.~B.}\ \bibnamefont {Kycia}}, \bibinfo {author}
  {\bibfnamefont {H.~A.}\ \bibnamefont {Dabkowska}}, \ and\ \bibinfo {author}
  {\bibfnamefont {B.~D.}\ \bibnamefont {Gaulin}},\ }\href@noop {} {\bibfield
  {journal} {\bibinfo  {journal} {arXiv:1312.0847 [cond-mat.stat-mech]}\ }
  (\bibinfo {year} {2013}{\natexlab{b}})}\BibitemShut {NoStop}%
\bibitem [{\citenamefont {Taniguchi}\ \emph {et~al.}(2013)\citenamefont
  {Taniguchi}, \citenamefont {Kadowaki}, \citenamefont {Takatsu}, \citenamefont
  {F\aa{}k}, \citenamefont {Ollivier}, \citenamefont {Yamazaki}, \citenamefont
  {Sato}, \citenamefont {Yoshizawa}, \citenamefont {Shimura}, \citenamefont
  {Sakakibara}, \citenamefont {Hong}, \citenamefont {Goto}, \citenamefont
  {Yaraskavitch},\ and\ \citenamefont {Kycia}}]{Taniguchi2013}%
  \BibitemOpen
  \bibfield  {author} {\bibinfo {author} {\bibfnamefont {T.}~\bibnamefont
  {Taniguchi}}, \bibinfo {author} {\bibfnamefont {H.}~\bibnamefont {Kadowaki}},
  \bibinfo {author} {\bibfnamefont {H.}~\bibnamefont {Takatsu}}, \bibinfo
  {author} {\bibfnamefont {B.}~\bibnamefont {F\aa{}k}}, \bibinfo {author}
  {\bibfnamefont {J.}~\bibnamefont {Ollivier}}, \bibinfo {author}
  {\bibfnamefont {T.}~\bibnamefont {Yamazaki}}, \bibinfo {author}
  {\bibfnamefont {T.~J.}\ \bibnamefont {Sato}}, \bibinfo {author}
  {\bibfnamefont {H.}~\bibnamefont {Yoshizawa}}, \bibinfo {author}
  {\bibfnamefont {Y.}~\bibnamefont {Shimura}}, \bibinfo {author} {\bibfnamefont
  {T.}~\bibnamefont {Sakakibara}}, \bibinfo {author} {\bibfnamefont
  {T.}~\bibnamefont {Hong}}, \bibinfo {author} {\bibfnamefont {K.}~\bibnamefont
  {Goto}}, \bibinfo {author} {\bibfnamefont {L.~R.}\ \bibnamefont
  {Yaraskavitch}}, \ and\ \bibinfo {author} {\bibfnamefont {J.~B.}\
  \bibnamefont {Kycia}},\ }\href {\doibase 10.1103/PhysRevB.87.060408}
  {\bibfield  {journal} {\bibinfo  {journal} {Phys. Rev. B}\ }\textbf {\bibinfo
  {volume} {87}},\ \bibinfo {pages} {060408} (\bibinfo {year}
  {2013})}\BibitemShut {NoStop}%
\bibitem [{\citenamefont {Molavian}\ \emph {et~al.}(2007)\citenamefont
  {Molavian}, \citenamefont {Gingras},\ and\ \citenamefont
  {Canals}}]{Molavian2007}%
  \BibitemOpen
  \bibfield  {author} {\bibinfo {author} {\bibfnamefont {H.~R.}\ \bibnamefont
  {Molavian}}, \bibinfo {author} {\bibfnamefont {M.~J.~P.}\ \bibnamefont
  {Gingras}}, \ and\ \bibinfo {author} {\bibfnamefont {B.}~\bibnamefont
  {Canals}},\ }\href@noop {} {\bibfield  {journal} {\bibinfo  {journal} {Phys.
  Rev. Lett.}\ }\textbf {\bibinfo {volume} {98}},\ \bibinfo {pages} {157204}
  (\bibinfo {year} {2007})}\BibitemShut {NoStop}%
\bibitem [{\citenamefont {Fennell}\ \emph {et~al.}(2009)\citenamefont
  {Fennell}, \citenamefont {Deen}, \citenamefont {Wildes}, \citenamefont
  {Schmalzl}, \citenamefont {Prabhakaran}, \citenamefont {Boothroyd},
  \citenamefont {Aldus}, \citenamefont {McMorrow},\ and\ \citenamefont
  {Bramwell}}]{Fennell2009}%
  \BibitemOpen
  \bibfield  {author} {\bibinfo {author} {\bibfnamefont {T.}~\bibnamefont
  {Fennell}}, \bibinfo {author} {\bibfnamefont {P.~P.}\ \bibnamefont {Deen}},
  \bibinfo {author} {\bibfnamefont {A.~R.}\ \bibnamefont {Wildes}}, \bibinfo
  {author} {\bibfnamefont {K.}~\bibnamefont {Schmalzl}}, \bibinfo {author}
  {\bibfnamefont {D.}~\bibnamefont {Prabhakaran}}, \bibinfo {author}
  {\bibfnamefont {A.~T.}\ \bibnamefont {Boothroyd}}, \bibinfo {author}
  {\bibfnamefont {R.~J.}\ \bibnamefont {Aldus}}, \bibinfo {author}
  {\bibfnamefont {D.~F.}\ \bibnamefont {McMorrow}}, \ and\ \bibinfo {author}
  {\bibfnamefont {S.~T.}\ \bibnamefont {Bramwell}},\ }\href@noop {} {\bibfield
  {journal} {\bibinfo  {journal} {Science}\ }\textbf {\bibinfo {volume}
  {326}},\ \bibinfo {pages} {415} (\bibinfo {year} {2009})}\BibitemShut
  {NoStop}%
\bibitem [{\citenamefont {Clancy}\ \emph {et~al.}(2009)\citenamefont {Clancy},
  \citenamefont {Ruff}, \citenamefont {Dunsiger}, \citenamefont {Zhao},
  \citenamefont {Dabkowska}, \citenamefont {Gardner}, \citenamefont {Qiu},
  \citenamefont {Copley}, \citenamefont {Jenkins},\ and\ \citenamefont
  {Gaulin}}]{Clancy2009}%
  \BibitemOpen
  \bibfield  {author} {\bibinfo {author} {\bibfnamefont {J. P.}~\bibnamefont
  {Clancy}}, \bibinfo {author} {\bibfnamefont {J. P. C.}~\bibnamefont {Ruff}},
  \bibinfo {author} {\bibfnamefont {S. R.}~\bibnamefont {Dunsiger}}, \bibinfo
  {author} {\bibfnamefont {Y.}~\bibnamefont {Zhao}}, \bibinfo {author}
  {\bibfnamefont {H. A.}~\bibnamefont {Dabkowska}}, \bibinfo {author}
  {\bibfnamefont {J. S.}~\bibnamefont {Gardner}}, \bibinfo {author} {\bibfnamefont
  {Y.}~\bibnamefont {Qiu}}, \bibinfo {author} {\bibfnamefont {J. R. D.}~\bibnamefont
  {Copley}}, \bibinfo {author} {\bibfnamefont {T.}~\bibnamefont {Jenkins}}, \
  and\ \bibinfo {author} {\bibfnamefont {B. D.}~\bibnamefont {Gaulin}},\
  }\href@noop {} {\bibfield  {journal} {\bibinfo  {journal} {Phys. Rev. B}\
  }\textbf {\bibinfo {volume} {79}},\ \bibinfo {pages} {014408} (\bibinfo
  {year} {2009})}\BibitemShut {NoStop}%
\bibitem [{\citenamefont {Hermele}\ \emph {et~al.}(2004)\citenamefont
  {Hermele}, \citenamefont {Fisher},\ and\ \citenamefont {Balents}}]{Hermele}%
  \BibitemOpen
  \bibfield  {author} {\bibinfo {author} {\bibfnamefont {M.}~\bibnamefont
  {Hermele}}, \bibinfo {author} {\bibfnamefont {M.~P.~A.}\ \bibnamefont
  {Fisher}}, \ and\ \bibinfo {author} {\bibfnamefont {L.}~\bibnamefont
  {Balents}},\ }\href {\doibase 10.1103/PhysRevB.69.064404} {\bibfield
  {journal} {\bibinfo  {journal} {Phys. Rev. B}\ }\textbf {\bibinfo {volume}
  {69}},\ \bibinfo {pages} {064404} (\bibinfo {year} {2004})}\BibitemShut
  {NoStop}%
\bibitem [{\citenamefont {Savary}\ and\ \citenamefont
  {Balents}(2013)}]{Savary}%
  \BibitemOpen
  \bibfield  {author} {\bibinfo {author} {\bibfnamefont {L.}~\bibnamefont
  {Savary}}\ and\ \bibinfo {author} {\bibfnamefont {L.}~\bibnamefont
  {Balents}},\ }\href {\doibase 10.1103/PhysRevB.87.205130} {\bibfield
  {journal} {\bibinfo  {journal} {Phys. Rev. B}\ }\textbf {\bibinfo {volume}
  {87}},\ \bibinfo {pages} {205130} (\bibinfo {year} {2013})}\BibitemShut
  {NoStop}%
\bibitem [{\citenamefont {Lee}\ \emph {et~al.}(2012)\citenamefont {Lee},
  \citenamefont {Onoda},\ and\ \citenamefont {Balents}}]{SBLee}%
  \BibitemOpen
  \bibfield  {author} {\bibinfo {author} {\bibfnamefont {S.~B.}\ \bibnamefont
  {Lee}}, \bibinfo {author} {\bibfnamefont {S.}~\bibnamefont {Onoda}}, \ and\
  \bibinfo {author} {\bibfnamefont {L.}~\bibnamefont {Balents}},\ }\href
  {http://link.aps.org/doi/10.1103/PhysRevB.86.104412} {\bibfield  {journal}
  {\bibinfo  {journal} {Phys. Rev. B}\ }\textbf {\bibinfo {volume} {86}},\
  \bibinfo {pages} {104412} (\bibinfo {year} {2012})}\BibitemShut {NoStop}%
\bibitem [{\citenamefont {Bonville}\ \emph {et~al.}(2011)\citenamefont
  {Bonville}, \citenamefont {Mirebeau}, \citenamefont {Gukasov}, \citenamefont
  {Petit},\ and\ \citenamefont {Robert}}]{Bonville2011}%
  \BibitemOpen
  \bibfield  {author} {\bibinfo {author} {\bibfnamefont {P.}~\bibnamefont
  {Bonville}}, \bibinfo {author} {\bibfnamefont {I.}~\bibnamefont {Mirebeau}},
  \bibinfo {author} {\bibfnamefont {A.}~\bibnamefont {Gukasov}}, \bibinfo
  {author} {\bibfnamefont {S.}~\bibnamefont {Petit}}, \ and\ \bibinfo {author}
  {\bibfnamefont {J.}~\bibnamefont {Robert}},\ }\href
  {http://link.aps.org/doi/10.1103/PhysRevB.84.184409} {\bibfield  {journal}
  {\bibinfo  {journal} {Phys. Rev. B}\ }\textbf {\bibinfo {volume} {84}},\
  \bibinfo {pages} {184409} (\bibinfo {year} {2011})}\BibitemShut {NoStop}%
\bibitem [{\citenamefont {Gaulin}\ \emph {et~al.}(2011)\citenamefont {Gaulin},
  \citenamefont {Gardner}, \citenamefont {McClarty},\ and\ \citenamefont
  {Gingras}}]{Gaulin2011}%
  \BibitemOpen
  \bibfield  {author} {\bibinfo {author} {\bibfnamefont {B.~D.}\ \bibnamefont
  {Gaulin}}, \bibinfo {author} {\bibfnamefont {J.~S.}\ \bibnamefont {Gardner}},
  \bibinfo {author} {\bibfnamefont {P.~A.}\ \bibnamefont {McClarty}}, \ and\
  \bibinfo {author} {\bibfnamefont {M.~J.~P.}\ \bibnamefont {Gingras}},\
  }\href@noop {} {\bibfield  {journal} {\bibinfo  {journal} {Phys. Rev. B}\
  }\textbf {\bibinfo {volume} {84}},\ \bibinfo {pages} {140402} (\bibinfo
  {year} {2011})}\BibitemShut {NoStop}%
\bibitem [{\citenamefont {Ruff}\ \emph {et~al.}(2007)\citenamefont {Ruff},
  \citenamefont {Gaulin}, \citenamefont {Castellan}, \citenamefont {Rule},
  \citenamefont {Clancy}, \citenamefont {Rodriguez},\ and\ \citenamefont
  {Dabkowska}}]{Ruff2007}%
  \BibitemOpen
  \bibfield  {author} {\bibinfo {author} {\bibfnamefont {J.~P.~C.}\
  \bibnamefont {Ruff}}, \bibinfo {author} {\bibfnamefont {B.~D.}\ \bibnamefont
  {Gaulin}}, \bibinfo {author} {\bibfnamefont {J.~P.}\ \bibnamefont
  {Castellan}}, \bibinfo {author} {\bibfnamefont {K.~C.}\ \bibnamefont {Rule}},
  \bibinfo {author} {\bibfnamefont {J.~P.}\ \bibnamefont {Clancy}}, \bibinfo
  {author} {\bibfnamefont {J.}~\bibnamefont {Rodriguez}}, \ and\ \bibinfo
  {author} {\bibfnamefont {H.~A.}\ \bibnamefont {Dabkowska}},\ }\href
  {http://link.aps.org/doi/10.1103/PhysRevLett.99.237202} {\bibfield  {journal}
  {\bibinfo  {journal} {Phys. Rev. Lett.}\ }\textbf {\bibinfo {volume} {99}},\
  \bibinfo {pages} {237202} (\bibinfo {year} {2007})}\BibitemShut {NoStop}%
\bibitem [{\citenamefont {Mamsurova}\ \emph {et~al.}(1986)\citenamefont
  {Mamsurova}, \citenamefont {Pigal'skii},\ and\ \citenamefont
  {Pukhov}}]{Mamsurova1986}%
  \BibitemOpen
  \bibfield  {author} {\bibinfo {author} {\bibfnamefont {L. G.}~\bibnamefont
  {Mamsurova}}, \bibinfo {author} {\bibfnamefont {K. S.}~\bibnamefont
  {Pigal'skii}}, \ and\ \bibinfo {author} {\bibfnamefont {K. H.}~\bibnamefont
  {Pukhov}},\ }\href@noop {} {\bibfield  {journal} {\bibinfo  {journal} {JETP
  Lett.}\ }\textbf {\bibinfo {volume} {43}},\ \bibinfo {pages} {755} (\bibinfo
  {year} {1986})}\BibitemShut {NoStop}%
\bibitem [{\citenamefont {Nakanishi}\ \emph {et~al.}(2011)\citenamefont
  {Nakanishi}, \citenamefont {Kumagai}, \citenamefont {Yoshizawa},
  \citenamefont {Matsuhira}, \citenamefont {Takagi},\ and\ \citenamefont
  {Hiroi}}]{Nakanishi2011}%
  \BibitemOpen
  \bibfield  {author} {\bibinfo {author} {\bibfnamefont {Y.}~\bibnamefont
  {Nakanishi}}, \bibinfo {author} {\bibfnamefont {T.}~\bibnamefont {Kumagai}},
  \bibinfo {author} {\bibfnamefont {M.}~\bibnamefont {Yoshizawa}}, \bibinfo
  {author} {\bibfnamefont {K.}~\bibnamefont {Matsuhira}}, \bibinfo {author}
  {\bibfnamefont {S.}~\bibnamefont {Takagi}}, \ and\ \bibinfo {author}
  {\bibfnamefont {Z.}~\bibnamefont {Hiroi}},\ }\href {\doibase
  10.1103/PhysRevB.83.184434} {\bibfield  {journal} {\bibinfo  {journal} {Phys.
  Rev. B}\ }\textbf {\bibinfo {volume} {83}},\ \bibinfo {pages} {184434}
  (\bibinfo {year} {2011})}\BibitemShut {NoStop}%
\bibitem [{\citenamefont {Ruff}\ \emph {et~al.}(2010)\citenamefont {Ruff},
  \citenamefont {Islam}, \citenamefont {Clancy}, \citenamefont {Ross},
  \citenamefont {Nojiri}, \citenamefont {Matsuda}, \citenamefont {Dabkowska},
  \citenamefont {Dabkowski},\ and\ \citenamefont {Gaulin}}]{Ruff2010}%
  \BibitemOpen
  \bibfield  {author} {\bibinfo {author} {\bibfnamefont {J.~P.~C.}\
  \bibnamefont {Ruff}}, \bibinfo {author} {\bibfnamefont {Z.}~\bibnamefont
  {Islam}}, \bibinfo {author} {\bibfnamefont {J.~P.}\ \bibnamefont {Clancy}},
  \bibinfo {author} {\bibfnamefont {K.~A.}\ \bibnamefont {Ross}}, \bibinfo
  {author} {\bibfnamefont {H.}~\bibnamefont {Nojiri}}, \bibinfo {author}
  {\bibfnamefont {Y.~H.}\ \bibnamefont {Matsuda}}, \bibinfo {author}
  {\bibfnamefont {H.~A.}\ \bibnamefont {Dabkowska}}, \bibinfo {author}
  {\bibfnamefont {A.~D.}\ \bibnamefont {Dabkowski}}, \ and\ \bibinfo {author}
  {\bibfnamefont {B.~D.}\ \bibnamefont {Gaulin}},\ }\href {\doibase
  10.1103/PhysRevLett.105.077203} {\bibfield  {journal} {\bibinfo  {journal}
  {Phys. Rev. Lett.}\ }\textbf {\bibinfo {volume} {105}},\ \bibinfo {pages}
  {077203} (\bibinfo {year} {2010})}\BibitemShut {NoStop}%
\bibitem [{\citenamefont {Dalmas~de R\'eotier}\ \emph
  {et~al.}(2006)\citenamefont {Dalmas~de R\'eotier}, \citenamefont {Yaouanc},
  \citenamefont {Keller}, \citenamefont {Cervellino}, \citenamefont {Roessli},
  \citenamefont {Baines}, \citenamefont {Forget}, \citenamefont {Vaju},
  \citenamefont {Gubbens}, \citenamefont {Amato},\ and\ \citenamefont
  {King}}]{Reotier2006}%
  \BibitemOpen
  \bibfield  {author} {\bibinfo {author} {\bibfnamefont {P.}~\bibnamefont
  {Dalmas~de R\'eotier}}, \bibinfo {author} {\bibfnamefont {A.}~\bibnamefont
  {Yaouanc}}, \bibinfo {author} {\bibfnamefont {L.}~\bibnamefont {Keller}},
  \bibinfo {author} {\bibfnamefont {A.}~\bibnamefont {Cervellino}}, \bibinfo
  {author} {\bibfnamefont {B.}~\bibnamefont {Roessli}}, \bibinfo {author}
  {\bibfnamefont {C.}~\bibnamefont {Baines}}, \bibinfo {author} {\bibfnamefont
  {A.}~\bibnamefont {Forget}}, \bibinfo {author} {\bibfnamefont
  {C.}~\bibnamefont {Vaju}}, \bibinfo {author} {\bibfnamefont {P.~C.~M.}\
  \bibnamefont {Gubbens}}, \bibinfo {author} {\bibfnamefont {A.}~\bibnamefont
  {Amato}}, \ and\ \bibinfo {author} {\bibfnamefont {P.~J.~C.}\ \bibnamefont
  {King}},\ }\href {\doibase 10.1103/PhysRevLett.96.127202} {\bibfield
  {journal} {\bibinfo  {journal} {Phys. Rev. Lett.}\ }\textbf {\bibinfo
  {volume} {96}},\ \bibinfo {pages} {127202} (\bibinfo {year}
  {2006})}\BibitemShut {NoStop}%
\bibitem [{\citenamefont {Rule}\ \emph {et~al.}(2009)\citenamefont {Rule},
  \citenamefont {Ehlers}, \citenamefont {Gardner}, \citenamefont {Qiu},
  \citenamefont {Moskvin}, \citenamefont {Kiefer},\ and\ \citenamefont
  {Gerischer}}]{Rule2009}%
  \BibitemOpen
  \bibfield  {author} {\bibinfo {author} {\bibfnamefont {K.~C.}\ \bibnamefont
  {Rule}}, \bibinfo {author} {\bibfnamefont {G.}~\bibnamefont {Ehlers}},
  \bibinfo {author} {\bibfnamefont {J.~S.}\ \bibnamefont {Gardner}}, \bibinfo
  {author} {\bibfnamefont {Y.}~\bibnamefont {Qiu}}, \bibinfo {author}
  {\bibfnamefont {E.}~\bibnamefont {Moskvin}}, \bibinfo {author} {\bibfnamefont
  {K.}~\bibnamefont {Kiefer}}, \ and\ \bibinfo {author} {\bibfnamefont
  {S.}~\bibnamefont {Gerischer}},\ }\href {\doibase
  10.1088/0953-8984/21/48/486005} {\bibfield  {journal} {\bibinfo  {journal}
  {J. Phys.: Condens. Matter}\ }\textbf {\bibinfo {volume} {21}},\ \bibinfo
  {pages} {486005} (\bibinfo {year} {2009})}\BibitemShut {NoStop}%
\bibitem [{\citenamefont {Bert}\ \emph {et~al.}(2006)\citenamefont {Bert},
  \citenamefont {Mendels}, \citenamefont {Olariu}, \citenamefont {Blanchard},
  \citenamefont {Collin}, \citenamefont {Amato}, \citenamefont {Baines},\ and\
  \citenamefont {Hillier}}]{Bert2006}%
  \BibitemOpen
  \bibfield  {author} {\bibinfo {author} {\bibfnamefont {F.}~\bibnamefont
  {Bert}}, \bibinfo {author} {\bibfnamefont {P.}~\bibnamefont {Mendels}},
  \bibinfo {author} {\bibfnamefont {A.}~\bibnamefont {Olariu}}, \bibinfo
  {author} {\bibfnamefont {N.}~\bibnamefont {Blanchard}}, \bibinfo {author}
  {\bibfnamefont {G.}~\bibnamefont {Collin}}, \bibinfo {author} {\bibfnamefont
  {A.}~\bibnamefont {Amato}}, \bibinfo {author} {\bibfnamefont
  {C.}~\bibnamefont {Baines}}, \ and\ \bibinfo {author} {\bibfnamefont {A.~D.}\
  \bibnamefont {Hillier}},\ }\href {\doibase 10.1103/PhysRevLett.97.117203}
  {\bibfield  {journal} {\bibinfo  {journal} {Phys. Rev. Lett.}\ }\textbf
  {\bibinfo {volume} {97}},\ \bibinfo {pages} {117203} (\bibinfo {year}
  {2006})}\BibitemShut {NoStop}%
\bibitem [{\citenamefont {Molavian}\ \emph {et~al.}(2009)\citenamefont
  {Molavian}, \citenamefont {McClarty},\ and\ \citenamefont
  {Gingras}}]{Molavian_arXiv}%
  \BibitemOpen
  \bibfield  {author} {\bibinfo {author} {\bibfnamefont {H.~R.}\ \bibnamefont
  {Molavian}}, \bibinfo {author} {\bibfnamefont {P.~A.}\ \bibnamefont
  {McClarty}}, \ and\ \bibinfo {author} {\bibfnamefont {M.~J.~P.}\ \bibnamefont
  {Gingras}},\ }\href@noop {} {\bibfield  {journal} {\bibinfo  {journal}
  {arXiv:0912.2957v1 [cond-mat.stat-mech]}\ } (\bibinfo {year}
  {2009})}\BibitemShut {NoStop}%
\bibitem [{\citenamefont {McClarty}\ \emph {et~al.}(2010)\citenamefont
  {McClarty}, \citenamefont {Stasiak},\ and\ \citenamefont
  {Gingras}}]{McClarty2010}%
  \BibitemOpen
  \bibfield  {author} {\bibinfo {author} {\bibfnamefont {P.~A.}\ \bibnamefont
  {McClarty}}, \bibinfo {author} {\bibfnamefont {P.}~\bibnamefont {Stasiak}}, \
  and\ \bibinfo {author} {\bibfnamefont {M.~J.~P.}\ \bibnamefont {Gingras}},\
  }\href@noop {} {\bibfield  {journal} {\bibinfo  {journal} {arxiv:1011.6346
  [cond-mat.stat-mech]}\ } (\bibinfo {year} {2010})}\BibitemShut {NoStop}%
\bibitem [{\citenamefont {Granroth}\ \emph {et~al.}(2010)\citenamefont
  {Granroth}, \citenamefont {Kolesnikov}, \citenamefont {Sherline},
  \citenamefont {Clancy}, \citenamefont {Ross}, \citenamefont {Ruff},
  \citenamefont {Gaulin},\ and\ \citenamefont {Nagler}}]{Granroth2010}%
  \BibitemOpen
  \bibfield  {author} {\bibinfo {author} {\bibfnamefont {G.~E.}\ \bibnamefont
  {Granroth}}, \bibinfo {author} {\bibfnamefont {A.~I.}\ \bibnamefont
  {Kolesnikov}}, \bibinfo {author} {\bibfnamefont {T.~E.}\ \bibnamefont
  {Sherline}}, \bibinfo {author} {\bibfnamefont {J.~P.}\ \bibnamefont
  {Clancy}}, \bibinfo {author} {\bibfnamefont {K.~A.}\ \bibnamefont {Ross}},
  \bibinfo {author} {\bibfnamefont {J.~P.~C.}\ \bibnamefont {Ruff}}, \bibinfo
  {author} {\bibfnamefont {B.~D.}\ \bibnamefont {Gaulin}}, \ and\ \bibinfo
  {author} {\bibfnamefont {S.~E.}\ \bibnamefont {Nagler}},\ }\href {\doibase
  10.1088/1742-6596/251/1/012058} {\bibfield  {journal} {\bibinfo  {journal}
  {J. Phys.: Conf. Series}\ }\textbf {\bibinfo {volume} {251}},\ \bibinfo
  {pages} {012058} (\bibinfo {year} {2010})}\BibitemShut {NoStop}%
\bibitem [{\citenamefont {Hutchings}(1964)}]{Hutchings}%
  \BibitemOpen
  \bibfield  {author} {\bibinfo {author} {\bibfnamefont {M.~T.}\ \bibnamefont
  {Hutchings}},\ }\href@noop {} {\emph {\bibinfo {title} {Solid State Physics,
  edited by F. Seitz and D. Thurnbull}}},\ Vol.~\bibinfo {volume} {16}\
  (\bibinfo  {publisher} {Academic},\ \bibinfo {address} {New York},\ \bibinfo
  {year} {1964})\ p.\ \bibinfo {pages} {227}\BibitemShut {NoStop}%
\bibitem [{\citenamefont {Kassman}(1970)}]{Kassman1970}%
  \BibitemOpen
  \bibfield  {author} {\bibinfo {author} {\bibfnamefont {A.~J.}\ \bibnamefont
  {Kassman}},\ }\href {\doibase 10.1063/1.1673904} {\bibfield  {journal}
  {\bibinfo  {journal} {J. Chem. Phys.}\ }\textbf {\bibinfo {volume} {53}},\
  \bibinfo {pages} {4118} (\bibinfo {year} {1970})}\BibitemShut {NoStop}%
\bibitem [{\citenamefont {Squires}(1978)}]{Squires1978}%
  \BibitemOpen
  \bibfield  {author} {\bibinfo {author} {\bibnamefont {Squires}},\ }\href@noop
  {} {\emph {\bibinfo {title} {Introduction to Thermal Neutron Scattering}}}\
  (\bibinfo  {publisher} {Cambridge University Press, Cambridge, UK},\ \bibinfo {year}
  {1978})\BibitemShut {NoStop}%
\bibitem [{\citenamefont {Kao}\ \emph {et~al.}(2003)\citenamefont {Kao},
  \citenamefont {Enjalran}, \citenamefont {Del~Maestro}, \citenamefont
  {Molavian},\ and\ \citenamefont {Gingras}}]{Kao2003}%
  \BibitemOpen
  \bibfield  {author} {\bibinfo {author} {\bibfnamefont {Y.-J.}\ \bibnamefont
  {Kao}}, \bibinfo {author} {\bibfnamefont {M.}~\bibnamefont {Enjalran}},
  \bibinfo {author} {\bibfnamefont {A.}~\bibnamefont {Del~Maestro}}, \bibinfo
  {author} {\bibfnamefont {H.~R.}\ \bibnamefont {Molavian}}, \ and\ \bibinfo
  {author} {\bibfnamefont {M.~J.~P.}\ \bibnamefont {Gingras}},\ }\href@noop {}
  {\bibfield  {journal} {\bibinfo  {journal} {Phys. Rev. B}\ }\textbf {\bibinfo
  {volume} {68}},\ \bibinfo {pages} {172407} (\bibinfo {year}
  {2003})}\BibitemShut {NoStop}%
\bibitem [{\citenamefont {Curnoe}(2013)}]{Curnoe2013}%
  \BibitemOpen
  \bibfield  {author} {\bibinfo {author} {\bibfnamefont {S.~H.}\ \bibnamefont
  {Curnoe}},\ }\href {\doibase 10.1103/PhysRevB.88.014429} {\bibfield
  {journal} {\bibinfo  {journal} {Phys. Rev. B}\ }\textbf {\bibinfo {volume}
  {88}},\ \bibinfo {pages} {014429} (\bibinfo {year} {2013})}\BibitemShut
  {NoStop}%
\bibitem [{Man()}]{Mantid}%
  \BibitemOpen
  \href@noop {} {\ }\bibinfo {note} {Http://www.mantidproject.org/,
  (2012)}\BibitemShut {NoStop}%
\bibitem [{\citenamefont {Azuah}\ \emph {et~al.}(2009)\citenamefont {Azuah},
  \citenamefont {Kneller}, \citenamefont {Qiu}, \citenamefont
  {Tregenna-Piggott}, \citenamefont {Brown}, \citenamefont {Copley},\ and\
  \citenamefont {Dimeo}}]{DAVE}%
  \BibitemOpen
  \bibfield  {author} {\bibinfo {author} {\bibfnamefont {R.~T.}\ \bibnamefont
  {Azuah}}, \bibinfo {author} {\bibfnamefont {L.~R.}\ \bibnamefont {Kneller}},
  \bibinfo {author} {\bibfnamefont {Y.}~\bibnamefont {Qiu}}, \bibinfo {author}
  {\bibfnamefont {P.~L.~W.}\ \bibnamefont {Tregenna-Piggott}}, \bibinfo
  {author} {\bibfnamefont {C.~M.}\ \bibnamefont {Brown}}, \bibinfo {author}
  {\bibfnamefont {J.~R.~D.}\ \bibnamefont {Copley}}, \ and\ \bibinfo {author}
  {\bibfnamefont {R.~M.}\ \bibnamefont {Dimeo}},\ }\href@noop {} {\bibfield
  {journal} {\bibinfo  {journal} {J. Res. Natl. Inst. Stan. Technol.}\ }\textbf
  {\bibinfo {volume} {114}},\ \bibinfo {pages} {341} (\bibinfo {year}
  {2009})}\BibitemShut {NoStop}%
\end{thebibliography}
\end{document}